\documentclass[aps,twocolumn,floats,nofootinbib,pre]{revtex4-2}
\usepackage{graphics,graphicx,epsfig}
\usepackage{amssymb,color}
\usepackage{amsmath}
\usepackage{epsf,epstopdf,wrapfig}
\usepackage[hidelinks]{hyperref}

\newcommand{\beq}{\begin{equation}}
\newcommand{\eeq}{\end{equation}}
\newcommand{\beqn}{\begin{eqnarray}}
\newcommand{\eeqn}{\end{eqnarray}}

\begin{document}

\title{Scale invariance in early embryonic development} 

\author{Milo\v{s} Nikoli\'c,$^{a,b}$ Victoria Antonetti,$^{a,c}$  Feng Liu,$^{b,c}$ Gentian Muhaxheri,$^{a,d}$\\
Mariela D.~Petkova,$^e$ Martin Scheeler,$^a$ Eric M. Smith,$^a$ William Bialek,$^{a,b,f}$ and Thomas Gregor$^{a,b,g}$}

\affiliation{$^a$Joseph Henry Laboratories of Physics and $^b$Lewis--Sigler Institute for Integrative Genomics, Princeton University, Princeton NJ 08544 USA\\
$^c$Center for Quantitative Biology and School of Physics, Peking University, Beijing 100871 China\\
$^d$Department of Physics, Lehman College, City University of New York, Bronx, NY 10468 USA\\
$^e$Program in Biophysics, Harvard University, Cambridge MA 02138 USA\\
$^f$Initiative for the Theoretical Sciences, The Graduate Center, City University of New York, 365 Fifth Ave., New York, NY 10016 USA\\
$^g$Department of Developmental and Stem Cell Biology UMR3738, Institut Pasteur, 75015 Paris, France}

\date{\today}

\begin{abstract} 
The body plan of the fruit fly is determined by the expression of just a handful of genes.  We show that the spatial patterns of expression for several of these genes scale precisely with the size of the embryo.  Concretely, discrete positional markers such as the peaks in striped patterns have absolute positions along the anterior--posterior axis that are proportional to embryo length, with better than $1\%$ accuracy.  Further, the information (in bits) that graded patterns of expression provide about position can be decomposed into information about fractional or scaled position and information about absolute position or embryo length; all of the available information is about scaled position, again with $\sim 1\%$ accuracy.  These observations suggest that the underlying genetic network exhibits scale invariance in a deeper mathematical sense.  Taking this mathematical statement seriously requires that the network dynamics have a zero mode, which connects to many other observations on this system.
\end{abstract}

\maketitle

\section{Introduction}

Closely related organisms can vary widely in size, but variations in their proportions are much smaller \cite{ishimatsu+al_18,almuedo-castillo+al_18,leibovich+al_20}.  There is a considerable gap between this qualitative observation and some precise mathematical statement of scaling, e.g., that the linear dimensions of all elements in the body plan are in direct proportion to the linear dimensions of the organism.  If correct this scale invariance would be visible not only in the fully developed organism but already at some earlier stages in its development.  

There are many examples of ``allometric scaling,'' power-law relationships among different quantities across a well-defined class of organisms \cite{huxley_32,mcmahon+bonner_1983,west+al_97}.  In some cases, these relations connect the linear dimensions of different body parts.  Nonetheless, truly precise spatial scaling in embryonic development would be quite surprising.  

We understand the mechanisms of pattern formation in a wide range of non--biological systems, from fluid flows to crystal growth (snowflakes) and more  \cite{kepler,benard_00,rayleigh_16,flesselles+al_91,cross+hohenberg_93,lappa_09}, but none of these examples exhibit scale invariance.  Instead, the elements of the pattern have linear dimensions set by microscopic parameters, and larger systems exhibit more repetitions of the same pattern rather than expansion or contraction of pattern elements to match the size of the system as a whole  \cite{langer_89}. Going back to the pioneering work of Turing \cite{turing_52}, the mathematical structure of the equations governing these systems is not so different from the structure of models for genetic or biochemical networks.  If we take these analogies literally, we would predict that taller people should have more vertebrae, which is obviously wrong.  Is there a real problem here, or are we just oversimplifying?

Here we try to make the notion of scale invariance in development more precise. We use the first hours of development in the fruit fly as an example, following spatial patterns of morphogen concentration as they flow through three layers of a genetic network, from maternal inputs to the gap genes to the pair rule genes \cite{scott+carroll_87,jaeger+verd_20,tkacik+gregor_21}. In the spirit of earlier work \cite{houchmandzadeh+al_02, surkova+al_08, holloway+al_06, lott+al_07,miles+al_11} we analyze discrete positional markers, such as the stripes in pair rule-gene expression, and find that positions of these markers vary in proportion to the length of the embryo with better than $1\%$ accuracy \cite{antonetti+al_18}.  We then go beyond discrete markers, decomposing the information carried by graded patterns of gap gene expression into information about fractional or scaled position vs.~information about the absolute position; we find that all the available information is about fractional position along the anterior--posterior axis. 
Information that would signal a deviation from scale invariance is less than $1\%$ of the total.

These results provide strong evidence for scaling in a precise mathematical sense, for both the gap genes and the pair rule genes.  But at least one of the maternal inputs, Bicoid \cite{driever+nusslein-volhard_88a,driever+nusslein-volhard_88b}, does not show any sign of scale invariance:  as in well-understood non-biological pattern-forming systems, there is a length scale that presumably is set by underlying molecular parameters and does not adjust in response to the linear dimensions of the whole embryo.  This suggests that scale invariance is an emergent property of the gap gene network.  

We argue that true scale invariance places very specific requirements on the dynamics of this network, independent of molecular details: it must have a ``zero mode.'' This has connections to other observations on gap gene dynamics \cite{krotov+al_14,mcgough+al_23} and to more detailed models \cite{manu+al_09b,vakulenko+al_09}.

\section{Testing for scaling}

To make the notion of scaling more precise we take seriously the idea that cell fates are determined by the concentration of particular molecules called morphogens \cite{wolpert_69}.  Since the cell fates are tied to their positions, the concentrations of morphogens must also carry information about position along the body axes.  These ideas are especially crisp in the early fly embryo, where we know the identities of all the relevant morphogens and rich spatial patterns in the concentrations of these molecules are established before cells make large-scale movements \cite{lawrence_92}.

We focus on pattern formation along a single axis, which will be the anterior--posterior axis in the analysis of fly embryos below.  Then we can measure position along this axis by a single variable $0 < x < L$, where $x=0$ is the anterior end of the embryo, $x=L$ is the posterior end, and hence $L$ is the length of the embryo.  
There are multiple morphogen species, indexed by $\rm i$, and if we neglect the discreteness of cells then their concentration profiles are described by continuous functions $g_{\rm i} (x; L)$.  The notation emphasizes that concentration profiles may be different in embryos of different size $L$.

True scale invariance is the statement that the concentration of morphogens depends only on position relative to the length, that is 
\begin{equation}
g_{\rm i} (x; L) = \Phi_{\rm i} (x/L).
\label{scale_def}
\end{equation}
If there is a fixed map from morphogen concentrations to cell fates, this scaling behavior would guarantee that cells adopt a fate that depends on their relative position $x/L$, and not separately on $x$ and $L$.

How do we test for scale invariance?  If the concentration of the morphogen has a single peak as a function of $x$, we can write
\begin{equation}
g_{\rm i} (x; L) = g (x - x_{\rm p}; L),
\end{equation}
then scale invariance as in Eq.~(\ref{scale_def}) requires that all the $L$ dependence is contained in the position of the peak.
\begin{equation}
x_{\rm p} = \langle f_{\rm p} \rangle \cdot L + {\rm noise};
\label{marker_scale}
\end{equation}
where $f_p$ is the fractional or scaled peak position, $\langle\cdots\rangle$ is the average over many embryos with different lengths, and $\rm noise$ allows that positions jitter from embryo to embryo.  We emphasize that Eq.~(\ref{marker_scale}) is not just the statement that positional markers adjust (in absolute distance) to the length of the embryo; scale invariance as we have defined it in Eq.~(\ref{scale_def}) requires that this adjustment is exactly linear with zero intercept.  There is a natural generalization to concentration profiles that have multiple peaks, as with the pair rule genes (Fig.~\ref{fig:1}A, B).

It has been known for some time that the morphogens in the early fly embryo carry enough information to specify scaled positions with $\sim 1\%$ precision all along the anterior--posterior axis \cite{dubuis+al_13a,petkova+al_19}.  At the same time, embryos from the same mother, in an inbred laboratory stock, fluctuate in length with a standard deviation of $\sigma_L / \langle L\rangle \sim 4\%$ (Appendix \ref{app:length} and \cite{gregor+al_07a,smith_15}).  It would seem that to make these numbers consistent with one another, positional signals must scale with embryo length, but this is a bit subtle.

Imagine a hypothetical embryo in which, e.g., the peak of the morphogen profile is perfectly anchored in absolute position relative to the anterior pole of the embryo, with no scaling and no noise,  such that $x_{\rm p} = \langle x_{\rm p}\rangle$. Then the relative or fractional positions $f_{\rm p} = x_{\rm p}/L$ fluctuate only because the lengths of the embryos vary,
\begin{eqnarray}
\sigma_{f_{\rm p}}^2 (A) &\equiv& \langle (\delta f_{\rm p})^2\rangle = \langle x_{\rm p}\rangle^2 \left[ \left\langle \left(\frac{1}{L}\right)^2\right\rangle - \left\langle \frac{1}{L}\right\rangle^2 \right] \\
&\sim& \left( {{ \langle x_{\rm p}\rangle}\over{\langle L\rangle}} \cdot {{\sigma_L }\over{\langle L\rangle}} \right)^2 .
\label{bound1}
\end{eqnarray}
Thus for a marker that on average is a quarter of the way from the anterior to posterior, $\langle x_{\rm p}\rangle = 0.25 \langle L\rangle$, fluctuations will be $\sigma_{f_{\rm p}} (A) \sim 0.01$ even without scaling.  Similarly, if we have a marker anchored at some fixed absolute position relative to the posterior then the variance in fractional position  will be
\begin{equation}
\sigma_{f_{\rm p}}^2 (P) = \left(1 - {{ \langle x_{\rm p}\rangle}\over{\langle L\rangle}}  \right)^2 \cdot \left(  {{\sigma_L }\over{\langle L\rangle}} \right)^2 .
\label{bound2}
\end{equation}
We can imagine cells combining anterior and posterior signals to reduce the error, 
\begin{equation}
{1\over {\sigma_{f_{\rm p}}^2 (A, P)}}  = {1\over {\sigma_{f_{\rm p}}^2 (A)}}  + {1\over {\sigma_{f_{\rm p}}^2 (P)}} .
\label{bound3}
\end{equation} 
With $\sigma_L /\langle L\rangle \sim 0.04$, fluctuations in fractional position thus could be less than $\sim 1.4\%$ everywhere along the anterior--posterior axis, even in the absence of any scaling mechanism.  Convincing ourselves that pattern formation is truly scale invariant requires a very precise measurement and depends on the system itself being very precise.

It is intuitive to think about scaling as the proportionality of positions to embryo length, as in Eq.~(\ref{marker_scale}), but it should be possible to test the scaling of the entire morphogen profile, as in Eq.~(\ref{scale_def}), more directly.  There are two related observations.  First, to compare morphogen profiles across embryos of different lengths, we need a metric.  Second, since morphogen profiles are noisy, it is unrealistic to expect the exact equality of two functions across all values of $x$.  Fortunately, the noise level itself provides a metric for comparison, which is made precise in the language of information theory.

The statement that morphogen profiles depend on $x$ and $L$ means that the concentrations of these molecules provide information about the underlying positional variables.  This is quantified, uniquely, by the Shannon information \cite{shannon_48,cover+thomas_91,bialek_12}, 
\begin{widetext}
\begin{equation}
I({\mathbf g} \rightarrow \{x , L\}) 
= \int  d{\mathbf g}  \int dx \int dL\, P\left( {\mathbf g} | \{x;L\}\right) P(x, L)  
\log_2\left[  {{P\left( {\mathbf g}  | \{x,L\}\right)}\over{P\left({\mathbf g} \right)}} \right]  \,{\rm bits},
\label{Ibits}
\end{equation}
\end{widetext}
where for more compact notation we write ${\mathbf g} = \{g_{\rm i}\}$ and $d{\mathbf g} = \prod_{\rm i} dg_{\rm i}$.
Here $P\left( {\mathbf g}  | \{x,L\}\right)$ is the probability of finding the set of morphogen concentrations $\{g_{\rm i}\} $ at position $x$ in an embryo of length $L$; $P\left( {\mathbf g} \right)$ is the probability of finding these concentrations averaged over all values of $x$ and $L$; and $P(x, L)$ is the distribution of positions and lengths.  It is useful to recall that this information is mutual: the concentrations of morphogens provide cells with information about position, and specifying position allows us to predict the concentrations, so we write $I({\mathbf g}  ; \{x , L\})$.   Information depends on both the mean spatial profiles of the morphogens and their noise levels.

True scale invariance would mean that all of the information conveyed by the morphogens is about the fractional position $x/L$:
\begin{equation}
I({\mathbf g} ;  \{x , L\})  = I({\mathbf g}  ; x / L) \,\,\,\, {\rm (perfect\ scaling)}.
\end{equation}
Equivalently, if we want to predict the morphogen concentration, it is enough to specify the fractional position, and no extra information is gained by knowing $x$ and $L$ separately.  We can think of the total information as having a component about the relative position and an extra increment that describes the deviation from scaling,
\begin{equation}
I( {\mathbf g}  ;  \{x , L\})  = I( {\mathbf g} ;  x / L) + \Delta I,
\label{I_decomp}
\end{equation}
and we will see that with samples from a sufficiently large number of embryos, we can make a reliable estimate of $\Delta I$.
The smaller the fraction $\Delta I/I( {\mathbf g}  ; x / L)$ the closer the system is to a mathematical ideal of scaling.  More explicit expressions for $\Delta I$ are developed in Appendix \ref{app:decomp} and applied to experiments in \S\ref{sec-info}.

We emphasize that true scale invariance, corresponding to $\Delta I = 0$, is a very strong condition.  Different levels of evidence for scaling in embryonic development have inspired models in which competing mechanisms can provide some cancellation of the intrinsic length scales determined by diffusion constants and reaction rates \cite{howard+wolde_05,houchmandzadeh+al_05,mchale+al_06,capek+muller_19}.  These models typically allow for scaling in the position of a single discrete positional marker (e.g., the middle of the embryo), or for approximate scaling across a larger segment of the relevant axes. True scale invariance would require new dynamical mechanisms.

\section{Stripes and boundaries}

\begin{figure}
\includegraphics[width = \linewidth]{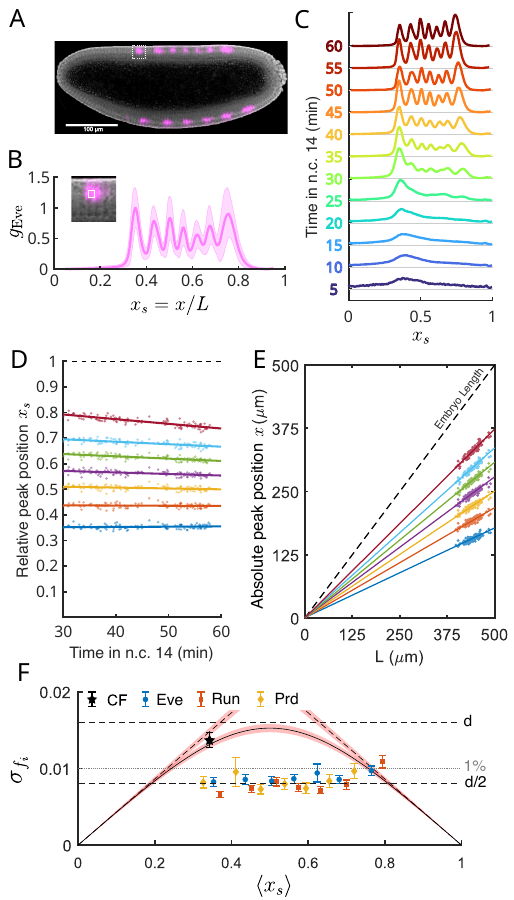} 
\caption{{\bf Precise scaling of pair-rule stripes in the {\em Drosophila} embryo.} (A) Bright-field image overlaid with fluorescent antibody staining for Eve protein (fuschia), focusing on the mid-sagittal plane with the dorsal side up; scalebar is $100\,\mu{\rm m}$. (B) Expression of Eve in the second half of nuclear cycle fourteen (nc14). Solid line is the mean, and shaded region is the standard deviation across $N_{\rm em} = 108$ embryos in a time window between 30 and $60\,$min from the start of nc14. Inset shows a single nucleus with a white square (width $0.01L$) used to average intensities. (C) Eve expression profiles as a function of relative position along the body axis for 12 time bins during nc14, as indicated by color.   (D) Linear dynamics of Eve peak positions during nc14, fit to Eq.~(\ref{linearA}).  (E) Absolute positions of Eve peaks measured from the anterior pole referred to $t_0 = 45\,{\rm min}$, as in Eq.~(\ref{linearB}),  plotted vs.~embryo length.   (F) Standard deviation of scaled stripe positions as a function of mean position for three pair-rule genes, and for the cephalic furrow (CF, see Appendix \ref{app:CF}).  Error bars are standard deviations from bootstrapping.    Black curves with red shading (bootstrapped errors) are estimates of precision based on anchoring in Eqs.~(\ref{bound1}--\ref{bound3}), and  $d$ is the spacing between neighboring cells. \label{fig:1} }
\end{figure}

In the early fly embryo, information about position along the anterior--posterior axis flows from maternal inputs through the network of gap genes to the pair-rule genes \cite{tkacik+gregor_21}.  The pair-rule genes are expressed in interdigitating striped patterns that provide a preview of the segmented body plan in the fully developed organism; these stripes are visible within three hours after the egg is laid (Fig.~\ref{fig:1}A--C).  The positions of pair-rule stripes are a clear example of the positional markers discussed above.

Here we analyze the spatial profiles of gene expression for three of the pair-rule genes---{\em eve}, {\em prd}, and {\em run}---measured using fluorescent antibody staining of the corresponding proteins in more than one hundred embryos that were fixed during nuclear cycle 14 (nc14), i.e. between 2 and $3\,$h of development \cite{petkova+al_19}.  Our results recapitulate earlier work \cite{antonetti+al_18} on a larger ensemble of embryos.

As soon as the stripes are visible it is straightforward to measure their positions $x_{\rm i}$ \cite{petkova+al_19}. The time during nc14 can be measured with $\sim 1\,{\rm min}$ precision by following the progression of the invaginating cellularization membrane \cite{dubuis+al_13a}.  The stripe positions vary systematically in time \cite{frasch+al_88, jaeger+al_04a, jaeger+al_04b, surkova+al_08, bothma+al_14}  and are well described by
\begin{equation}
\frac{x_{\rm i}(t)}{L} = \frac{x_{\rm i}(t_0)}{L}+ s_{\rm i} (t-t_0) ,
\label{linearA}
\end{equation}
as shown for the Eve stripes in Fig~\ref{fig:1}D.  Combining data from all time points, we shift each embryo to the reference time $t_0 = 45\,{\rm min}$,
\begin{equation}
\frac{x_{\rm i}(t)}{L} \rightarrow \frac{x_{\rm i}(t)}{L} -  s_{\rm i} (t-t_0) .
\label{linearB}
\end{equation}
We use this same procedure for the Prd and Run stripes, although these become clear only at slightly later times.

Figure \ref{fig:1}E shows that the stripe positions $x_{\rm i}$ measured from the anterior pole are proportional to the length of the embryo $L$. More precisely, if we fit these linear relations then intercepts are zero and slopes are equal to the mean fractional positions, as in Eq.~(\ref{marker_scale}), both results with error bars of less than $1\%$ (Appendix \ref{app:CF}). This provides \emph{prima facie} evidence for scaling of the pair-rule stripes, reinforcing the conclusions of earlier work  \cite{houchmandzadeh+al_02, surkova+al_08, holloway+al_06, lott+al_07}.

We can go beyond the mean behaviors to look at fluctuations around these means. For each stripe $\rm i$ in each embryo $\alpha$, we can write
\begin{equation}
{{x_{\rm i}^\alpha}\over{L^\alpha}} = \langle f_{\rm i} \rangle + \delta f_{\rm i}^\alpha ,
\end{equation}
where $\langle \cdots \rangle$ now is an average over all the embryos in our sample.  The variance of the relative position is $\sigma_{f_{\rm i}}^2 = \langle (\delta f_{\rm i})^2\rangle$, and Fig.~\ref{fig:1}F shows that  $\sigma_{f_{\rm i}} \leq 0.01$ for all 21 pair rule stripes that we measure.  This is consistent with previous measurements, and with the information content of the gap gene expression patterns that feed into the generation of pair-rule stripes \cite{dubuis+al_13b, petkova+al_19}, but earlier work did not address scaling explicitly.

As a caution, we note that the observation of scaling in fixed embryos would be trivial if variations in embryo length were dominated by shrinkage during fixation.  Across $N_{\rm em} = 609$ fixed embryos used for the analysis of gap genes (below) we find a mean length $\langle L \rangle_{\rm fix} = 455\,\mu{\rm m}$, while across $N_{\rm em} = 610$ live embryos (\S\ref{maternal}) we find $\langle L \rangle_{\rm live} = 490\,\mu{\rm m}$. Hence, shrinkage with fixation is a bit less than $10\%$ across many different experiments.  But the variations in  length are almost the same, $(\sigma_L/\langle L\rangle)_{\rm fix} = 0.038$, while $(\sigma_L/\langle L\rangle)_{\rm live} = 0.037$.  The small extra variance in the length of fixed embryos cannot explain the scaling behavior that we observe.

Figure \ref{fig:1}F also shows that the fluctuations in fractional position are smaller than the bound on mechanisms that have no explicit scaling, from Eq.~(\ref{bound3}). This bound is very tight, because of the small variance in emrbyo lengths, and thus requires extreme precision in the measurement and biological reproducibility of the fractional positions to demonstrate scaling. To emphasize the importance of precision, we note that the position of the cephalic furrow is directly regulated by pair rule gene expression \cite{vincent+al_97}, but it has a slightly higher relative positional variance,  due to the experimental difficulty of defining morphological features to less than the width of a single cell \cite{liu+al_13}.  Here we show explicitly that the furrow position scales with embryo length (Appendix \ref{app:CF}). Even though the precision of the CF position is almost $\sim 1\%$ in the scaled coordinates \cite{liu+al_13}, this alone is not sufficient to reject the hypothesis that positions are defined in absolute rather than relative coordinates, as can be seen from Fig.~\ref{fig:1}F.

\begin{figure}[t]
\includegraphics[width = \linewidth]{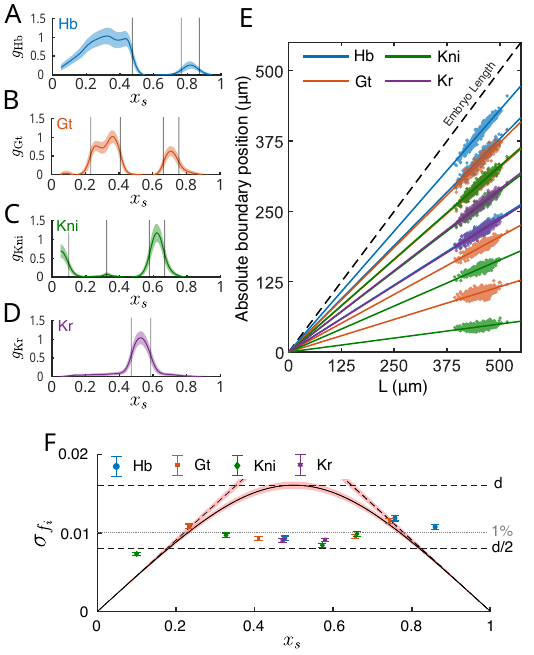}
\caption{\textbf{Precise scaling of gap gene expression boundaries.} Expression profiles of  (A) Hunchback (Hb), (B) Giant (Gt), (C) Knirps (Kni), and (D) Kr\"uppel (Kr), based on immunofluorescent staining (Appendix \ref{app:gapdata}). Means (solid lines) and standard deviations (shading) across embryos aligned by scaled position $x_s$. Vertical lines indicate the mean positions of expression boundaries as well as a small peak in Kni.
(E) Absolute position of all gap gene boundaries as a function of the embryo length. Dashed black line indicates the position of the posterior of the embryo. Boundary positions are time-corrected to $t_0 = 45\,{\rm min}$, as with the stripe positions in Fig.~\ref{fig:1}D.
(F) Standard deviation of scaled boundary positions as a function of mean position for all 13 markers.  Error bars are standard deviations from bootstrapping.    Black curves with red shading (bootstrapped errors) are estimates of precision based on anchoring in Eqs.~(\ref{bound1}--\ref{bound3}), and  $d$ is the spacing between neighboring cells.
Horizontal dashed lines denote the distance $d$ and half-distance $d/2$, between neighboring nuclei. Dotted gray line indicates 1\% precision. 
\label{fig:2}}
\end{figure}

The pair rule stripes are shaped by input from the gap genes \cite{jaeger_11}, and it is natural to ask whether the scaling behavior that we observe is inherited from these inputs.     The gap genes were long discussed in terms of ``expression domains,'' as if they were on/off switches \cite{kauffman+al_78,meinhardt_86,albert+othmer_03,spirov+holloway_86}. 
We now know that this misses a substantial fraction of the positional information encoded by these genes \cite{dubuis+al_13b,tkacik+al_15,petkova+al_19}, but defining the boundaries of the expression domains as positional markers (Fig.~\ref{fig:2}A--D) allows us to give a preliminary analysis of scaling by following the same ideas as for the positions of the pair-rule stripes.  

Previous experiments have measured the expression profiles of the gap genes \cite{petkova+al_19}, staining $N_{\rm em} = 609$ fixed embryos in nc14 with fluorescent antibodies directed at the proteins encoded by the gap genes (Fig.~\ref{fig:2}A--D).  We define expression boundaries as the positions where the concentrations are half their maximum mean value, and we correct their relative positions to $t_0=45$ min as above.   Figure \ref{fig:2}E shows that all thirteen of the gap gene boundaries defined in this way have absolute positions that scale precisely with embryo length, as with the positions of the pair rule stripes.  The accuracy of this scaling again is better than $\sim 1\%$, and this precision is better than the limiting performance of mechanisms that do not have some explicit sensitivity to embryo length (Fig.~\ref{fig:2}F).
For the gap genes, this procedure allows us to span almost the entire range of the anterior--posterior axis.  

In summary, stripes and boundaries of gene expression in the early fly embryo provide discrete positional markers, and the absolute positions of these markers are in all cases proportional to the length of the embryo.  This is consistent with previous observations \cite{houchmandzadeh+al_02, surkova+al_08, holloway+al_06, lott+al_07}, but the precision of the scaling that we observe here is surprising.  This suggests that the underlying genetic network exhibits true scale invariance, which we now test using the information decomposition [Eq.~(\ref{I_decomp})].

\section{Absolute vs.~scaled positional information}
\label{sec-info}

The concentrations of morphogens provide cells with information about their position in the embryo.  This ``positional information'' \cite{wolpert_69} can be measured in bits if we have access to data on the mean and variability of spatial profiles for the concentration of the relevant molecules \cite{dubuis+al_13b,tkacik+al_15}.  The local expression levels of individual gaps genes convey roughly two bits of information about position, twice what is possible in a model of on/off expression domains. Taken together all four gap genes provide $\sim 4.2\,{\rm bits}$, sufficient to specify positions with $\sim 1\%$ accuracy along the anterior--posterior axis, as seen above.  However, these earlier analyses assumed, implicitly, that information is about the fractional or scaled position.  Is this correct?  

\begin{figure}[b]
\centering{\includegraphics[width = 0.9\linewidth]{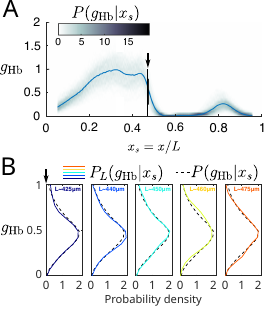}}
\caption{{\bf Expression of Hb in scaled coordinates.}  (A) Mean concentration of Hb, $\langle g_{\rm Hb} (x_s)\rangle$, vs scaled position (solid line, as in Fig.~\ref{fig:2}A) and the conditional distribution $P(g_{\rm Hb}|x_s)$ around this mean (shading). Intensity bin size is 0.05 maximum $\langle g_{\rm Hb} \rangle$.   (B) A slice through the conditional distribution at $x_s= \ 0.47$ (dashed black lines) compared with distributions estimated from embryos in narrow bins of length, $P_L(g_{\rm Hb}| x_s)$.   Lengths were binned in 5 bins with an equal number of embryos in each, such that each bin contains about 60 embryos with variations in $L$ of less than $1\%$. Mean lengths in each bin are indicated at the upper right of each panel. Probability distributions of $g_{\rm Hb}$ are estimated using a kernel density estimator with a Gaussian kernel that has width $\delta g = 0.07 \times \max_{x_s}\langle g_{\rm Hb} (x_s)\rangle$. 
\label{fig:Hb}}
\end{figure}

\begin{figure*}[t]
\centerline{\includegraphics[width = \linewidth]{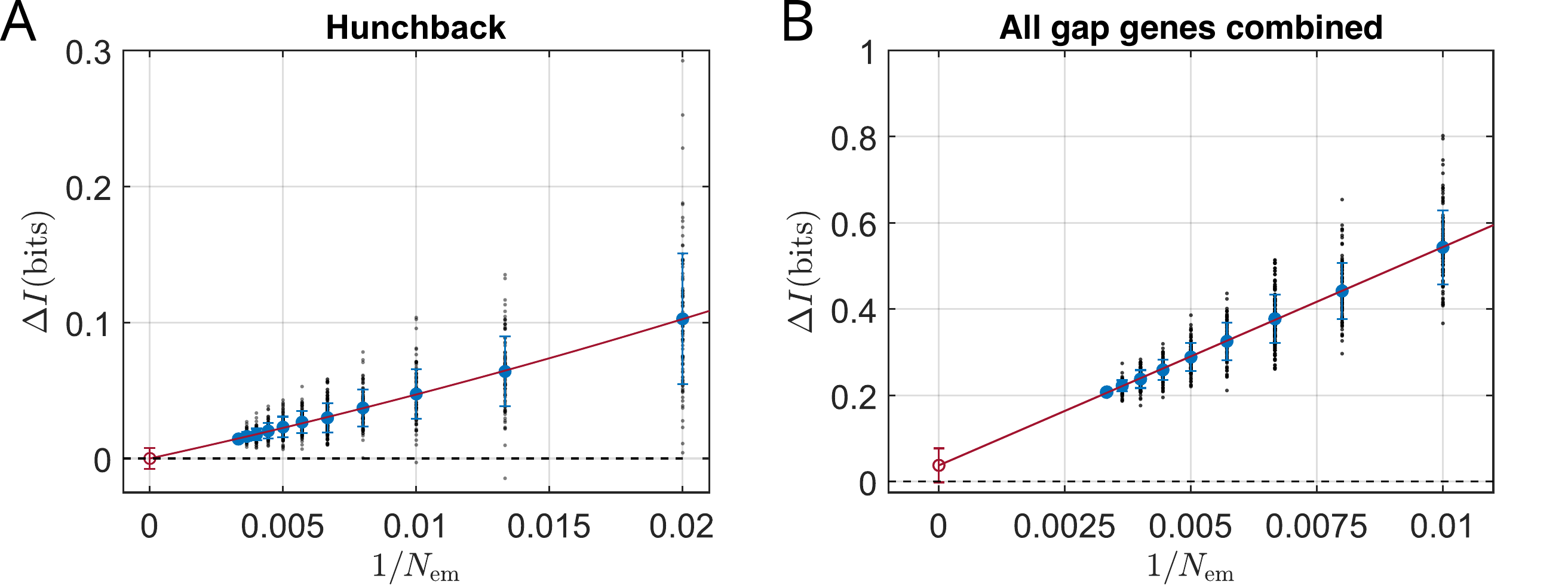}}
\caption{{\bf Near zero deviation from perfect scaling, in bits.} (A) The extra information $\Delta I$ that Hb expression levels carry about absolute rather than scaled position, defined by Eq.~(\ref{I_decomp}) and evaluated from Eq.~(\ref{DeltaI_vars-main}).  Estimates are based on random choices of $N_{\rm }$ embryos out of the full experimental ensemble (points; circles show means with standard deviations), and the extrapolation $N_{\rm em}\rightarrow \infty$ follows the methods of Appendix \ref{app:finitedata} (red line). The result is $\Delta I = 0.00 \pm 0.008\,{\rm bits}$ (red circle with error bar). (B) The extra information $\Delta I$ conveyed by all four gap genes together, defined as in (A) by Eq.~(\ref{I_decomp}) but now evaluated using Eq.~(\ref{DI-multigene-main}).  Symbols as in (A); the result is $\Delta I = 0.038 \pm 0.039\,{\rm bits}$.  Error bars are larger because we are analyzing a multidimensional code, but there still is no significant difference from $\Delta I = 0$.
 \label{fig:Igap}}
\end{figure*}

The key to separating information about scaled vs.~absolute position is to compare the variance in morphogen concentrations at a scaled position $x_s$ depending on whether we constrain the length of the embryo (Appendix \ref{app:decomp}).  Qualitatively, if there is perfect scaling then knowing the length would not add any information with which to predict the morphogen concentration. Since information is mutual this would mean that all the available information is about the scaled position.  To test this quantitatively in the context of the gap genes, we have assembled data on $N_{\rm em} = 301$ embryos, in each of which we have reliable simultaneous measurements on the spatial profiles of expression in all four gap genes, as described in Appendix \ref{app:gapdata}.

Figure \ref{fig:Hb}A shows the spatial profile of Hb as a function of scaled position along the anterior--posterior axis.  At each scaled position $x_s = x/L$ we can visualize the distribution of expression levels, which is well approximated as a Gaussian (Appendix \ref{app:gapdata} and \cite{dubuis+al_13b}).  We can then ask if this distribution changes when we look only at embryos in a narrow range of lengths $L$, and the answer is no (qualitatively; Fig.~\ref{fig:Hb}B).  Quantitatively we want to estimate the difference in entropy between these two distributions, averaged over $x_s$ and $L$, which will give us the deviation from scaling $\Delta I$ in Eq.~(\ref{I_decomp}), as explained in Appendix \ref{app:decomp}.  The calculation of the entropy simplifies in the Gaussian approximation, depending just on the variances as in Eq.~(\ref{DeltaI_vars}),
\begin{equation}
\Delta I = {1\over 2}\langle\log_2[ \sigma_g^2(x_s) ]\rangle_{x_s} -  {1\over 2}\langle \log_2[ \sigma_g^2(x_s|L)]\rangle_{x_s,L},
\label{DeltaI_vars-main}
\end{equation}
where $\sigma_g^2(x_s|L)$ is the variance in concentration at scaled position $x_s$ across embryos of length $L$ and  $\sigma_g^2(x_s)$ is the same variance computed across all embryos.

Applying Eq.~(\ref{DeltaI_vars-main}) requires estimating the relevant variances and also making bins along the $x_s$ and $L$ axes.  For the scaled position we choose bins of size $\Delta x_s = 0.01$, consistent with the precision that we see in Figs.~\ref{fig:1} and \ref{fig:2}.  To sample the range of embryo lengths we use $N_{\rm bins} = 5,\, 10,\, 15,$ or $20$ adaptive bins, and find the same results in all cases (Appendix \ref{app:finitedata}).  As is well known, estimates of entropy or information are subject to systematic errors \cite{miller_55,bialek_12}.  In the present case, if we substitute estimates of the variances into Eq.~(\ref{DeltaI_vars-main}), we find a nonzero result for $\Delta I$.  But suppose we include different numbers of embryos in our analysis. In that case, we see that our estimate of $\Delta I$ depends on $1/N_{\rm em}$ as expected theoretically \cite{miller_55,bialek_12}, and having seen this predicted dependence we can extrapolate $N_{\rm em}\rightarrow\infty$. In particular, if we shuffle the data so that the true $\Delta I=0$, then our estimation procedure returns a random number with zero mean and standard deviation equal to our quoted error bar, demonstrating that we have control over the systematic errors.  These now standard analysis methods are explained more fully in Appendix \ref{app:finitedata}.  

Results of this analysis for Hb are shown in Fig.~\ref{fig:Igap}A.  Using all $N_{\rm em} = 301$ embryos in our data set produces a very small estimate of $\Delta I$, but even this is exaggerated by systematic errors as we see by changing $N_{\rm em}$.  Our best estimate extrapolates to zero as $N_{\rm em} \rightarrow\infty$, with an error bar smaller than $0.01\,{\rm bits}$.  When we repeat the same analyses for each of the other gap genes (i.e., Gt, Kni, and Kr), we get the same result (Appendix \ref{app:finitedata}).

\begin{figure*}[t]
\includegraphics[width = \linewidth]{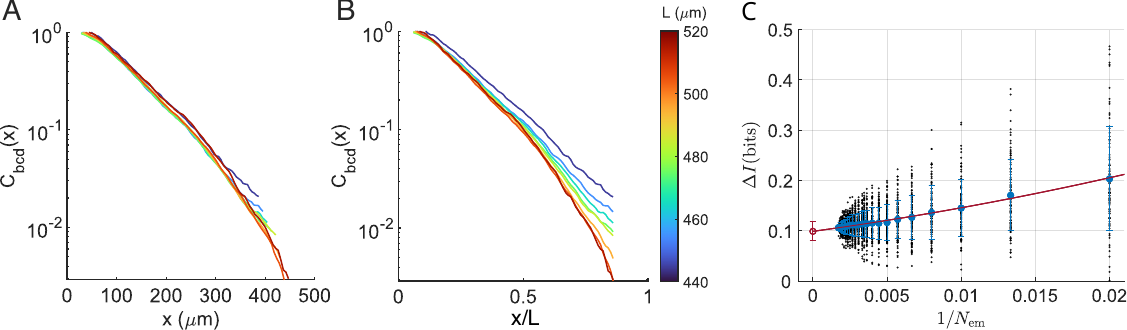}
\caption{{\bf The maternal input Bicoid does not scale.} (A) Measurements of Bcd concentration in $N_{\rm em} = 582$ live embryos are grouped into eight classes by embryo length $L$ and averaged.  There is only one global normalization, so this shows that absolute concentrations have the same dependence on absolute position $x$ across all classes. (B) The same data plotted vs.~scaled position $x_s = x/L$. Profiles separate, providing evidence against scaling. (C) Extra information $\Delta I$ that Bcd concentration provides about absolute vs.~scaled position, defined by Eq.~(\ref{I_decomp}) and evaluated from Eq.~(\ref{DeltaI_vars-main}). Symbols as in Fig.~\ref{fig:Hb}, but the extrapolation now leads to a significantly nonzero value of $\Delta I = 0.1\pm 0.02\,{\rm bits}$. Data from \cite{liu+al_13}. 
\label{fig:3} }
\end{figure*} 

We can generalize this analysis to consider all four gap genes simultaneously.  Now the role of the variance in Eq.~(\ref{DeltaI_vars-main}) is played by the covariance matrix of the fluctuations, as in Eq.~(\ref{DI-multigene}):
\begin{equation}
\Delta I = {1\over 2}\langle\log_2\left[ ||\Sigma(x_s)|| \right]\rangle_{x_s} -  {1\over 2}\langle \log_2 \left[||\Sigma(x_s|L)|| \right]\rangle_{x_s,L} .
\label{DI-multigene-main}
\end{equation}
Here $||\Sigma(x_s|L)||$ is the determinant of the covariance matrix describing fluctuations in the expression levels of all four genes at scaled position $x_s$ across embryos of length $L$, and  $\Sigma(x_s|L)$ is the covariance computed across all embryos.  Because we are looking at higher dimensional variations the impact of the finiteness of our data set is larger, but again we see the predicted dependence on $1/N_{\rm em}$ and can extrapolate to give $\Delta I = 0.038\pm 0.039\,{\rm bits}$ (Fig.~\ref{fig:Igap}B).  Once again this is consistent with $\Delta I = 0$: there is no significant evidence for encoding of information about absolute, as opposed to scaled position. 

Although the number of bits has meaning, it is useful to express the  deviation from perfect scaling as a fraction of the information available about scaled position \cite{dubuis+al_13b,tkacik+al_15},
\begin{equation}
{{I( {\mathbf g}  \rightarrow \{x, L\} ) - I ( {\mathbf g} \rightarrow x/L)}\over{I( {\mathbf g}  \rightarrow x/L)}} = 0.009 \pm 0.009.
\end{equation}
Thus the patterns of gap gene expression scale with $1\%$ accuracy, not just at discrete positional markers but across the entire range of graded spatial variations.

\section{Maternal inputs do not scale}
\label{maternal}

Having observed scaling in the pair rule stripe positions and followed this back to the gap genes, it is natural to ask if we can go one step further and trace the scaling behavior of the patterning system to the maternal inputs. Of the three maternal inputs that drive patterning along the anterior--posterior axis of the fly embryo, much attention has been given to Bicoid (Bcd) \cite{driever+nusslein-volhard_88a,driever+nusslein-volhard_88b}.  The protein is present at high concentrations in the anterior, and there is a nearly exponential decay of concentration with distance toward the posterior; one can monitor the dynamics of Bicoid protein concentrations quantitatively in live embryos using GFP-fusions \cite{gregor+al_07a}.  

Comparison across closely related species of flies shows that the length scale of this exponential decay varies in proportion to the mean length of the embryo \cite{gregor+al_05}.  Insertion of {\em bicoid} genes from other species into {\em Drosophila melanogaster} produces protein concentration profiles with length scales appropriate to the host, but these are not sufficient to rescue the embryo from deletion of the native Bcd \cite{gregor+al_08}.  These results emphasize the subtlety of comparison across species and the impact of genetic variations, leading us to re-examine the behavior of Bcd profiles across a large number of live embryos drawn from the same inbred laboratory strain used in the analysis of gap and pair rule genes.

Figure \ref{fig:3} analyzes Bcd profiles from $N_{\rm em} = 582$ live embryos \cite{liu+al_13}.  Measurements are taken during a small temporal window in nuclear cycle fourteen  \cite{gregor+al_07a}, and the only normalization (as with the gap genes) is to subtract a common background level from all the embryos and set the highest mean concentration to one.  When we group the embryos into eight classes based on their length $L$, we see that the average concentration profiles in all groups are the same when plotted vs.~absolute position, except for small effects at the posterior pole (Fig.~\ref{fig:3}A).  If we plot vs.~scaled position the different groups of embryos separate significantly (Fig.~\ref{fig:3}B), providing direct evidence {\em against} scaling.  We make this precise using the same information theoretic approach as above and now find a significant nonzero value of $\Delta I = 0.1\pm 0.02\,{\rm bits}$ (Fig.~\ref{fig:3}C).  This may seem like a small number, but this is related to the $\sim 4\%$ scale of variations in embryo length.  We conclude that the maternal inputs do not scale, in agreement with earlier suggestions \cite{houchmandzadeh+al_02}.

We emphasize that the absence of scaling in the maternal inputs should not be interpreted as a form of noise. Indeed, absolute concentrations of Bcd protein are highly reproducible across embryos and this can be traced to highly reproducible numbers of mRNA molecules \cite{gregor+al_07b,liu+al_13,petkova+al_14}.  Instead, we should think of the maternal inputs as a nearly deterministic response to the boundary conditions in the embryo, which also have a direct impact on the gap genes; see Eqs.~(\ref{noflux}, \ref{bcdflux}) below.

 \section{Scaling and zero modes}
 
The results here strongly support the view that patterns of gap gene expression are genuinely scale invariant and that this is an emergent property of the gap gene network.  Here we take the precise mathematical notion of scale invariance literally and explore its implications.  While we do not pretend to have a detailed model, it is useful to have in mind a  class of models for how patterns might form.  As a caution we recall Turing's introductory remarks \cite{turing_52}: ``This model will be a simplification and an idealization, and consequently a falsification.''

If we focus on variations just along the anterior--posterior axis $x$, and ignore  the discreteness of nuclei, then the concentration $g_{\rm i}$ of protein encoded by gene $\rm i$ plausibly obeys an equation of the form
\begin{equation}
{{\partial g_{\rm i}}\over{\partial t}} = D_{\rm i} {{\partial^2 g_{\rm i}}\over{\partial x^2}} + R_{\rm i} F_{\rm i} ({\mathbf g} ) - {1\over{\tau_{\rm i}}} g_{\rm i} .
\label{eq_dyn}
\end{equation}
Here $D_{\rm i}$ is the diffusion constant of species $\rm i$, $R_{\rm i} $ is the maximum rate at which these proteins can be synthesized, $\tau_{\rm i}$ is their lifetime before being degraded, and $F_{\rm i} ({\mathbf g} )$ describes all the potentially complex interactions by which all the proteins regulate the expression of gene $\rm i$.  We assume that the mRNA and protein dynamics have separate time scales so that one is effectively slaved to the other and we can write only one variable for each gene.  Further, we neglect time scales that might arise in the process of regulation itself, such as switching between different regulatory states, so that $F_{\rm i} ({\mathbf g} )$ is an instantaneous function of the relevant concentrations.  These assumptions are quite conventional, and other than this what we have written is very general.  For example, the function $F_{\rm i} ({\mathbf g} )$ could describe both activating and repressive interactions, and these interactions could be combinatorial.  These equations include as special cases Turing's original models \cite{turing_52} and their intellectual descendants \cite{gierer+meinhardt_72,meinhardt_08}.

The maximum steady state concentration of each protein is $R_{\rm i}\tau_{\rm i}$, and we can choose units in which this is equal to one, as with the normalized profiles of gap gene expression in Fig.~\ref{fig:2}A--D.  For simplicity we will assume that all the decay times are the same, $\tau_{\rm i} = \tau$, although this is not essential for what follows; finally, we choose units of time such that $\tau = 1$. Then we have
\begin{equation}
 {{\partial g_{\rm i}}\over{\partial t}} = \lambda_{\rm i}^2 {{\partial^2 g_{\rm i}}\over{\partial x^2}} +  F_{\rm i} ({\mathbf g} ) - g_{\rm i} ,
\label{eq_dyn2}
\end{equation}
where the length scale $\lambda_{\rm i} = \sqrt{D_{\rm i}\tau}$.  This describes an autonomous network, which is not quite realistic for the gap genes---which are driven by maternal inputs---but should be sufficient to draw qualitative conclusions about the implications of scale invariance.

The length of the embryo appears not in the dynamical equations but in the boundary conditions.  For most proteins, there can be no diffusive flux of molecules into or out of the ends of the embryo, so that
\begin{equation}
- D_{\rm i} {{\partial g_{\rm i}}\over{\partial x}}{\bigg |}_{x=0} = D_{\rm i} {{\partial g_{\rm i}}\over{\partial x}}{\bigg |}_{x=L} =0.
\label{noflux}
\end{equation}
The situation for maternal inputs is different;  as an example, in making the egg the mother places mRNA for the protein Bicoid (Bcd) at the anterior end ($x=0$), and this is translated continuously, so that
\begin{equation}
- D_{\rm Bcd} {{\partial g_{\rm Bcd}}\over{\partial x}}{\bigg |}_{x=0} = T_{\rm Bcd} ,
\label{bcdflux}
\end{equation}
where $T_{\rm Bcd}$ is the rate of translation in appropriate units.

Let us imagine that the final pattern we observe is in steady state, so that 
\begin{equation}
0 = \lambda_{\rm i}^2 {{\partial^2 g_{\rm i}(x;L)}\over{\partial x^2}} +  F_{\rm i} ({\mathbf g} ) - g_{\rm i} (x;L),
\label{eq_ss}
\end{equation}
where the notation reminds us that length dependence can arise once we impose the boundary conditions.  If we have true scale invariance as in Eq.~(\ref{scale_def}) then if we make a small change in the length of the embryo, so that $L\rightarrow L +\delta L$, the expression levels should change as
\begin{eqnarray}
g_{\rm i}(x;L) &\rightarrow& g_{\rm i}(x;L) + {{\delta L}\over{L}}\psi_{\rm i} (x/L)\\
\psi_{\rm i} (x_s ) &=& - x_s  \Phi_{\rm i}'(x_s) ,
\end{eqnarray}
but Eq.~(\ref{eq_ss}) still must be true.  This requires that 
\begin{equation}
\sum_{\rm j}\left[ \left( \lambda_{\rm i}^2 {{\partial^2 }\over{\partial x^2}}- 1\right)\delta_{\rm ij} + {{\partial F_{\rm i}}\over {\partial g_{\rm j}}}{\bigg |}_{{\mathbf g} = {\mathbf \Phi}} \right] \psi_{\rm j} (x/L) = 0 .
\label{zm1}
\end{equation}
This seemingly abstract condition has a direct implication for the dynamics of the network.

Suppose that the system is close to its steady state so that we can write 
\begin{equation}
g_{\rm i} (x;L ; t) = \Phi_{\rm i} (x/L) + \delta g_{\rm i} (x ; t) 
\end{equation}
and $\delta {\mathbf g}$ is small.   Then we can linearize the dynamics in Eq.~(\ref{eq_dyn2}) to yield
\begin{equation}
{{\partial (\delta g_{\rm i})}\over{\partial t}} = \sum_{\rm j}\left[ \left( \lambda_{\rm i}^2 {{\partial^2 }\over{\partial x^2}}- 1\right)\delta_{\rm ij} + {{\partial F_{\rm i}}\over {\partial g_{\rm j}}}{\bigg |}_{{\mathbf g} = {\mathbf \Phi}} \right] \delta g_{\rm j} .
\label{lin_dyn}
\end{equation}
We recognize the term in brackets as the same one that appears in Eq.~(\ref{zm1}).  To understand this connection it is useful to think of all possible spatial patterns of gene expression as points in a high dimensional space.  

Concretely we can write
\begin{equation}
\delta g_{\rm i} (x; t) = \sum_\mu a_\mu (t) \phi_{\rm i}^\mu (x) 
\end{equation}
where the functions $\{\phi_{\rm i}^\mu (x)\}$ 
are the spatial ``modes'' of expression and the set $\{a_\mu\}$ provides the coordinates of one expression profile in this multidimensional space.  The number of modes is the number of genes times the number of independent points along the $x$ axis, e.g. the number of rows of cells; for the gap genes the result is that the space has a dimensionality $d > 300$.  We can choose these modes as eigenfunctions of the operator that appears in both Eqs.~(\ref{zm1}) and (\ref{lin_dyn}),
\begin{equation}
\sum_{\rm j}\left[ \left( \lambda_{\rm i}^2 {{\partial^2 }\over{\partial x^2}}- 1\right)\delta_{\rm ij} + {{\partial F_{\rm i}}\over {\partial g_{\rm j}}}{\bigg |}_{{\mathbf g} = {\mathbf \Phi}} \right] \phi_{\rm j}^\mu (x) = - \lambda_\mu \phi_{\rm i}^\mu (x),
\end{equation}
where $\lambda_\mu \geq 0$ means that the steady state is stable.  Then so long as the deviations from the steady state are small, the dynamics of the network are simple in this coordinate system,
\begin{equation}
{ {da_\mu (t) }\over{dt}} = -\lambda_\mu a_\mu (t) .
\end{equation}
Through Eq.~(\ref{zm1}) we see that perfect scale invariance implies a ``zero mode,'' a particular mode of gene expression associated with eigenvalue $\lambda_\mu = 0$.  Importantly this is not the steady state pattern itself, but an additional mode.

The existence of a zero mode has several implications:
\begin{itemize}
\item Most literally, one component in the spatial pattern of gene expression will relax very slowly to its steady state, much more slowly than other components. Formally the relaxation should be as a power of time rather than exponential.
\item The dynamics describe a ``restoring force'' that pulls the patterns of gene expression toward their steady state values; the eigenvalues are the spring constants associated with these restoring forces.  Along the zero mode, there is no (linear) restoring force, and in the presence of any finite noise, the fluctuations along this mode will be very large compared with other modes.
\item Along directions with nonzero $\lambda_\mu$  the fluctuations in $\mathbf g$ will be approximately Gaussian so long as they remain small, as we see for the gap genes.  But along the zero mode, there should be some deviation from Gaussian behavior.
\end{itemize}
 There is evidence that the spatial patterns of gap gene expression can be compressed into a lower dimensional space, consistent with the idea that a zero mode dominates the dynamics \cite{seyboldt+al_22}.  The ($4\times4$) covariance matrix of fluctuations in gap gene expression is dominated by a single mode at almost all locations along the anterior--posterior axis, this large variance mode relaxes $\sim 10\times$ more slowly than the lower variance modes, and one can even see hints of non--Gaussian behavior \cite{krotov+al_14}.
 
The existence of a zero mode is a statement about the linearized dynamics.  If the absence of a linear restoring force continues for finite deviations from the steady state then there is a line of attracting spatial patterns rather than a single stable pattern.  Different points along this line are the patterns appropriate to embryos of different lengths, and the final pattern is selected by boundary conditions.  Line attractors have long been discussed for neural networks \cite{seung_96}.  It has been noted that models of the gap gene network might support such line attractors \cite{manu+al_09b}, and there are also suggestions that internal dynamics of the network can generate approximate scaling \cite{vakulenko+al_09}.  The observation of nearly perfect scale invariance in the real network leads us to a much sharper version of these ideas.

\section{Discussion}
 
Scale invariance is an appealing concept.  It quantifies the intuition that organisms are built from parts that are in proportion to one another, independent of an individual organism's overall size.  There is a long history of searching for such scaling not just in adult organisms but at early stages of development, and the fruit fly {\em Drosophila melanogaster} has been a particular target for these studies \cite{houchmandzadeh+al_05,holloway+al_06,lott+al_07,surkova+al_08,vakulenko+al_09}. If we compare related species of flies we can see spatial patterns of gene expression that scale, on average, across $10\times$ changes in embryo length \cite{gregor+al_05,gregor+al_08}, and similar results are obtained within a single species but with artificial selection for length variation \cite{miles+al_11}.  It has always been less clear whether scaling occurs without such large genetic variations, across the natural length variations in a single species.

We have explored scaling across many embryos from a quasi-inbred laboratory stock, minimizing genetic variation.  Across this ensemble, we see length fluctuations with a standard deviation of $\pm 4\%$ but embryos in the tails of the distribution have lengths $\pm 10\%$ from the mean (Fig.~\ref{fig:Ldist}). Following previous work, we measured the positions of discrete markers---such as the CF position, the peaks of pair-rule stripes, and the boundaries of gap gene domains---and found precise scaling of the absolute positions with embryo length.  This is consistent with previous results, but what is new is the precision that we observe:  markers are at positions that are scaled relative to the embryo length with an accuracy of $\sim 1\%$ across the full extent of the anterior--posterior axis. This observed precision excludes a broad class of models that combine information from both ends of the embryo without explicit scaling \cite{howard+wolde_05,houchmandzadeh+al_05,mchale+al_06,capek+muller_19}. 

There remains a gap between the positioning of discrete markers and the fuller notion of scale invariance.  The gap gets smaller as we track more markers across a wider range of positions, but it would be attractive to address scale invariance directly.  
We have introduced an information theoretic approach that analyzes the full, graded spatial profiles of gene expression and measures similarity in the natural units provided by the intrinsic noise levels of these profiles.  Concretely, we introduce a decomposition of the information that morphogen concentrations provide about position into a component about scaled position and a deviation from scaling.  Applied to the gap genes in the early fly embryo, the result is clear: the deviation from scaling is less than one percent of the total positional information. It is perhaps surprising that we can make such a precise statement about the functional output of a complex network.

In contrast to the results for the gap genes and the pair-rule genes, at least one of the maternal inputs, Bicoid, does not exhibit scaling.  We can see this ``by eye,'' simply plotting profiles vs.~absolute or scaled position, and these impressions are quantified by the same information theoretic approaches used to demonstrate scaling in the gap genes.  Error bars again are in the range of $\sim 0.01\,{\rm bits}$, but the deviation from scaling now is $\sim 10\times$ as large.  The conclusion is that near-perfect scale invariance is an emergent property of the gap gene network.

If we take scale invariance as a precise mathematical statement then the dynamics of the underlying genetic network must have a zero mode.  This is equivalent to saying that the dynamics do not have a single attractor, but rather a line of attractors as in models for short-term memory in neural networks \cite{seung_96}. Then position along this line is chosen by the boundary conditions and hence the length of the embryo.  A zero mode would provide connections among several otherwise disparate observations on the gap genes.  

Finally, recent experiments on mammalian pseudo-embryos suggest that scale invariance may be a more universal feature of genetic networks underlying developmental pattern formation \cite{merle+al_23}. In these self-organizing cell aggregates derived from stem cells, scale invariance emerges without fixed boundary conditions, but rather with boundaries that move as the aggregate grows. The existence of a zero mode in the regulatory network becomes even more attractive as a general mechanism for scaling.

\begin{acknowledgments}
We are grateful to E.~F. Wieschaus for his advice and for many inspiring discussions.  We thank  M.~Biggin and N.~Patel for sharing the antibodies used in these measurements. This work was supported in part by US National Science Foundation Grant PHY–1734030 (Center for the Physics of Biological Function); by National Institutes of Health Grants R01GM077599 and R01GM097275; by the Simons Foundation; by the John Simon Guggenheim Memorial Foundation.
\end{acknowledgments}

\renewcommand{\thefigure}{A\arabic{figure}} 
\setcounter{figure}{0}
\appendix

\section{Natural length variations of embryos in a laboratory strain of flies}
\label{app:length}

As described in the main text, much previous work on scaling has exploited the natural variation in embryo lengths across the evolutionary tree or the variations that can be selected artificially over reasonable times.  Here we use variations in length that occur within a single laboratory strain, \textit{OreR}, minimizing genetic variations.  Measurements on large numbers of live embryos are taken from Refs.~\cite{liu+al_13,smith_15} and on fixed embryos from Ref.~\cite{petkova+al_19}.

\begin{figure}[h]
\includegraphics[width=\linewidth]{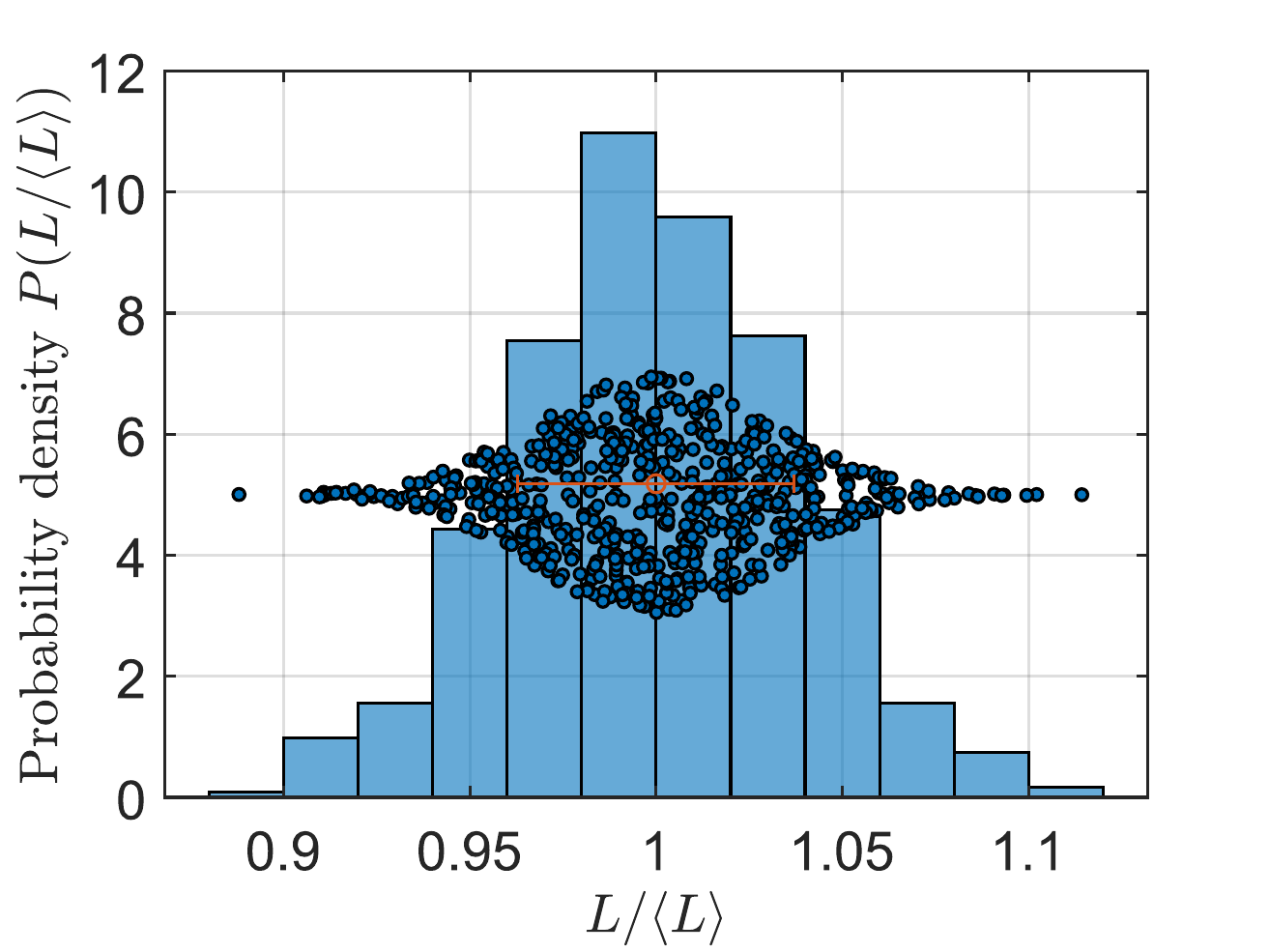}
\caption{{\bf Distribution of live embryo lengths.} Data from $N_{\rm em} = 610$ embryos \cite{liu+al_13} analyzed in bins of size $\Delta L/\langle L \rangle = 0.02$. The mean embryo length $\langle L \rangle = 490\pm 0.76\,\mu{\rm m}$. Overlaid is the error bar indicating the standard deviation of $\sigma_L = 0.0371\langle L\rangle$, and each dot indicates the length of one of the embryos in our sample. 
\label{fig:Ldist}}
\end{figure}

As an example, Fig.~\ref{fig:Ldist} shows the probability distribution of embryo lengths $L$ estimated from $N_{\rm em} = 610$ living dechorionated embryos (Bcd-GFP 2XA strain in \cite{liu+al_13}).  The mean length of the embryos is $\langle L \rangle = 490\pm 0.76\,\mu{\rm m}$, and the standard deviation is $\sigma_L = 18\pm 1.06\,\mu{\rm m}$.  This corresponds to a fractional variation $\sigma_L/\langle L\rangle = 0.037$, and as noted in the main text our sample is sufficiently large that it includes embryos $\pm10\%$ from the mean. This is true also in the case of fixed \textit{OreR} embryos where we find $\sigma_L/\langle L\rangle = 0.038$ and $\sigma_L/\langle L\rangle = 0.039$ in the experimental ensembles used for the analysis fo the gap and pair rule genes, respectively.

\section{Decomposing information}
\label{app:decomp}

We want to give explicit expressions that allow us to decompose positional information, as in Eq.~(\ref{I_decomp}), based on estimates from real data.  We give more detail than usual in hopes of making the analysis accessible to a broader audience.  

Concentrations can depend both on the absolute position $x$ and the length of the embryo $L$.  We can rewrite this as a dependence on the scaled position $x_s = x/L$ and the length $L$.  Thus we have
\begin{equation}
P \left({\mathbf g} |\{x,L\} \right) = P \left( {\mathbf g} |\{x/L , L\} \right) = P_L ( \{g_{\rm i}\}|x_s ),
\end{equation}
where $P_L $ is a distribution formed only across embryos of length $L$.  Similarly, we expect that cells are uniformly distributed along the $x$ axis up to the length $L$, so that
\begin{equation}
P(x,L) = {1\over L} \Theta(1-x_s) P(L),
\end{equation}
where $\Theta$ is the unit step function.  Then we can substitute into Eq.~(\ref{Ibits}):
\begin{widetext}
\begin{eqnarray}
I(\{g_{\rm i}\} \rightarrow \{x , L\}) 
&=& \int d{\mathbf g}  \int dx \int dL\, P\left( {\mathbf g}  | \{x;L\}\right) P(x, L)  
\log_2\left[  {{P\left( {\mathbf g}  | \{x,L\}\right)}\over{P\left( {\mathbf g}  \right)}} \right]  \\
&=& \int dL\, P(L) \int_0^1 dx_s \,P_L ( {\mathbf g}  |x_s) \log_2\left[  {{P_L ( {\mathbf g} |x_s)}\over{P\left( {\mathbf g} \right)}} \right]  .
\end{eqnarray}
Now we insert a factor of unity:
\begin{eqnarray}
\log_2\left[  {{P_L ( {\mathbf g} |x_s )}\over{P\left( {\mathbf g} \right)}} \right] &=&
\log_2\left[  
{{P_L ( {\mathbf g} |x_s )}\over{P\left( {\mathbf g} \right)}} 
{{P ({\mathbf g} |x_s )}\over{P ( {\mathbf g} |x_s )}} 
\right] \\
&=&
\log_2
\left[  
{ {P ( {\mathbf g} |x_s )} \over {P\left( {\mathbf g} \right)} }
\right]
-\log_2\left[P ( {\mathbf g} |x_s )\right] + \log_2\left[P_L ({\mathbf g} |x_s )\right] .
\end{eqnarray}
Substituting, we can write
\begin{equation}
I({\mathbf g}  \rightarrow \{x , L\})  = I_1 + I_2 + I_3,
\end{equation}
where the three components are
\begin{eqnarray}
I_1 &=& \int_0^1 dx_s \int d{\mathbf g} \, P({\mathbf g} |x_s) \log_2\left[  
{{P ( {\mathbf g} |x_s )}\over{P\left( {\mathbf g} \right)}} \right]\\
I_2 &=& - \int_0^1 dx_s \int  d{\mathbf g}  \,P({\mathbf g} |x_s )\log_2 [P ( {\mathbf g} |x_s )] \\
I_3 &=& \int dL \,P(L)\int_0^1 dx_s \int   d{\mathbf g}  \,P_L ( {\mathbf g} |x_s )\log_2 [P_L ({\mathbf g} |x_s )],
\end{eqnarray}
\end{widetext}

We can identify the three terms:  First, $I_1$ is the information that the concentrations of morphogens provide about the scaled position,
\begin{equation}
I_1 = I({\mathbf g}  \rightarrow x/L) .
\end{equation}
Second, $I_2$ is the entropy of the distribution of concentrations at a particular value of scaled position $x_s$, averaged over this position,
\begin{equation}
I_2 = \langle S[P({\mathbf g} |x_s)] \rangle_{x_s} ,
\end{equation}
where $S[Q]$ denotes the entropy of the distribution $Q$.  Finally, $I_3$ is the negative of the entropy of the same distribution but restricted to embryos of length $L$, and then averaged also over $L$,
\begin{equation}
I_3 = -\langle S[P_L({\mathbf g} |x_s)] \rangle_{x_s,L} .
\end{equation}
Comparing with Eq.~(\ref{I_decomp}) we see that the deviation from scaling can be written as the difference between two entropies, suitably averaged:
\begin{equation}
\Delta I = \langle S[P({\mathbf g} |x_s)] \rangle_{x_s}-\langle S[P_L({\mathbf g} |x_s)] \rangle_{x_s,L} .
\label{deltaIdef2}
\end{equation}
This has a very simple interpretation:  There is a deviation from scaling if specifying the length of the embryo reduces the entropy of fluctuations in morphogen concentration at a given scaled position.

These general expressions simplify enormously in the case where we have only a single morphogen and the conditional distributions are Gaussian.  In this case
\begin{eqnarray}
P(g |x_s) &=& {1\over {Z(x_s)}} \exp\left[ -{1\over 2}\chi^2 (g; x_s)\right]\\
\chi^2 (g; x_s) &=& {{[g - \langle g(x_s)\rangle]^2}\over{\sigma_g^2(x_s)}}\\
Z(x_s) &=& \sqrt{2\pi \sigma_g^2(x_s)} ,
\end{eqnarray}
where $\langle g(x_s)\rangle$ is the mean and $\sigma_g^2(x_s)$ is the variance of $g$ at scaled positions $x_s$.  Importantly the entropy of a Gaussian distribution does not depend on the mean, and we have \cite{cover+thomas_91,bialek_12}
\begin{equation}
S\left[P(g |x_s)\right] = {1\over 2}\log_2 \left[2\pi e \sigma_g^2(x_s)\right] .
\end{equation}
Thus we find the deviation from scaling is
\begin{equation}
\Delta I = {1\over 2}\langle\log_2[ \sigma_g^2(x_s) ]\rangle_{x_s} -  {1\over 2}\langle \log_2[ \sigma_g^2(x_s|L)]\rangle_{x_s,L},
\label{DeltaI_vars}
\end{equation}
where $\sigma_g^2(x_s|L)$ is the variance in concentration at scaled position $x_s$ across embryos of length $L$.  In  other words, there is a deviation from scaling if the variance in morphogen concentration is reduced by knowing the length of the embryo.

This result can be generalized to multiple morphogens if we stay in the Gaussian approximation.  Now with $d$ genes, we have
\begin{widetext}
\begin{eqnarray}
P({\mathbf g} |x_s) &=& {1\over {Z(x_s)}} \exp\left[ -{1\over 2}\chi^2 (\{g_{\rm i} \}; x_s)\right]\\
\chi^2 (\{g_{\rm i} \}; x_s) &=& \sum_{{\rm i}=1}^d \sum_{{\rm j}=1}^d [g_{\rm i} - \langle g_{\rm i}(x_s)\rangle] [\Sigma^{-1}(x_s)]_{\rm ij} [g_{\rm j} - \langle g_{\rm j}(x_s)\rangle] \\
Z(x_s) &=& \left[ (2\pi)^d ||\Sigma (x_s)||\right]^{1/2},
\end{eqnarray}
\end{widetext}
where $\Sigma (x_s)$ is the covariance matrix of fluctuations in concentration at scaled position $x_s$ and $||\Sigma (x_s)||$ is the determinant of this matrix.  Following the same logic as in the case of one gene we have
\begin{equation}
\Delta I = {1\over 2}\langle\log_2\left[ ||\Sigma(x_s)|| \right]\rangle_{x_s} -  {1\over 2}\langle \log_2 \left[||\Sigma(x_s|L)|| \right]\rangle_{x_s,L} .
\label{DI-multigene}
\end{equation}

Even if $P_L({\mathbf g}|x_s)$ is perfectly Gaussian, averaging over $L$ could generate non-Gaussian behavior in $P({\mathbf g}|x_s)$.  We are neglecting this here, but since we find that $\Delta I$ is very small, and for the gap genes consistent with $\Delta I = 0$, both the conditional and averaged distributions are very nearly Gaussian, as seen in Fig.~\ref{fig:Igap}B.

\section{Cephalic furrow and scale invariance}
\label{app:CF}

Upon the onset of gastrulation (i.e., three hours after fertilization), the cephalic furrow (CF) emerges as the first macroscopic morphological feature along the anterior--posterior axis in the developing fly embryo. It results from collective cell movement that can be seen using bright-field microscopy.  There are hints in early experiments that this marker is positioned very precisely \cite{driever+nusslein-volhard_88b}.  Modern experiments show that, as a fraction of the embryo length $L$, CF position $x_{\rm CF}$ is reproducible to nearly $1\%$ accuracy \cite{liu+al_13}.  

\begin{figure}
\centering{\includegraphics[width=0.8\linewidth]{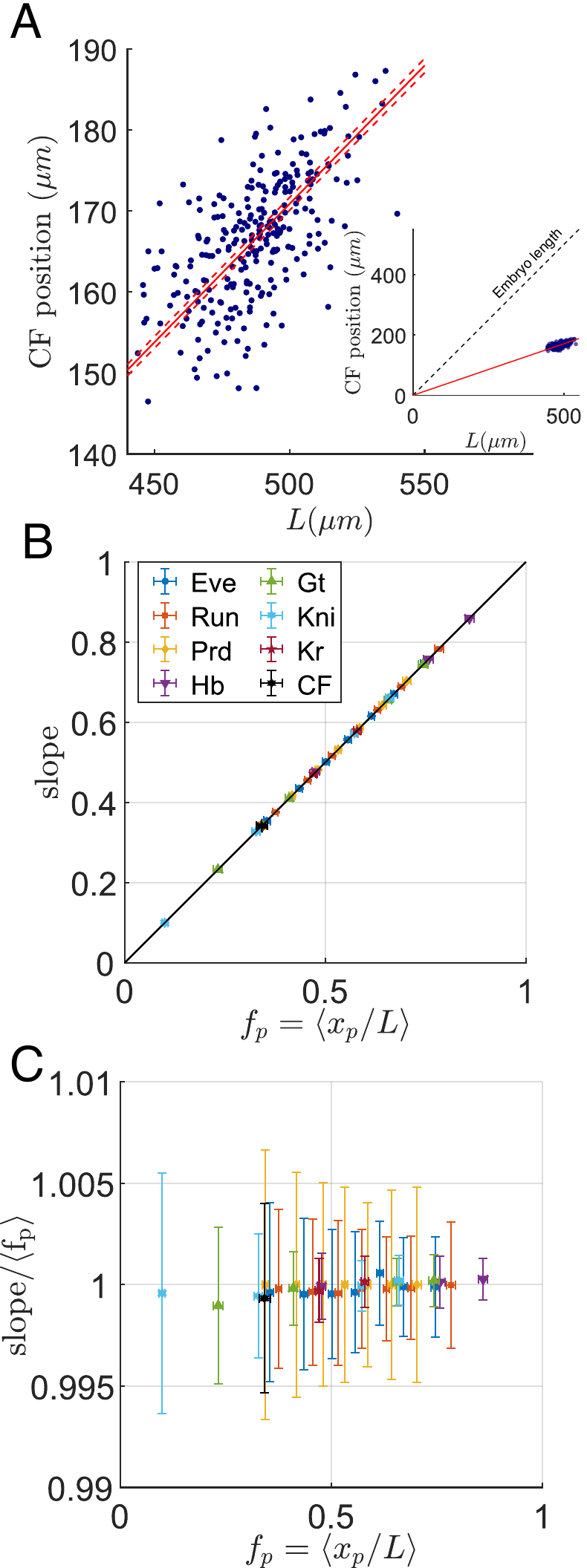}}
\caption{{\bf Cephalic furrow and the proportionality of scaling.} (A) The absolute position of the cephalic furrow measured in live emrbyos \cite{liu+al_13} is proportional to the embryo length. Red line is the best fit with 95\% confidence intervals shown as dashed lines. The entire fit is shown in the inset, emphasizing that the intercept is zero. 
(B) Slopes of absolute position vs.~embryo length for multiple positional markers, each plotted vs.~its mean scaled position.  (C) Replotting of data in (B) to show that slopes and scaled positions are equal within $1\%$, as predicted for perfect scaling [Eq.~(\ref{marker_scale})]. 
\label{fig:CFscale+slopes}}
\end{figure}

When we plot CF position in absolute units as a function of embryo length we observe a linear relationship with zero intercept (Fig.~\ref{fig:CFscale+slopes}A). The slope of this fit is well within 1\% of the mean scaled position $\langle f_{\rm CF} \rangle$.  More generally, all the discrete positional markers that we track (CF, pair-rule stripes, gap boundaries) have absolute positions that vary linearly with embryo length; the intercepts of the best-fit linear relations are zero; the slopes match the mean scaled positions of the markers  (Fig.~\ref{fig:CFscale+slopes}B) as predicted by scale invariance [Eq.~(\ref{marker_scale})]; and the precision of this match is better than $1\%$.

\section{Aspects of the gene expression data}
\label{app:gapdata}

We analyze gene expression patterns for the pair-rule genes, the gap genes, and the maternal input Bicoid. In each case, the concentration of the protein is inferred from the intensity of a fluorescence signal.  In each case images are collected by focusing on the midsagittal plane, the extent of the embryo is defined by thresholding the fluorescence intensity, and to avoid geometric distortions we avoid the $5\%$ of the embryo at both the anterior and posterior poles. Fluorescence intensities are averaged over a small area, as shown in the inset to Fig.~\ref{fig:1}B, sliding along the dorsal rim of the embryo. In live embryos, we can keep track of time during nc14 directly, while in fixed embryos we use the progress of the cellularization as a clock with precision $\sim 1\,{\rm min}$.

For each gene $\rm i$ we measure an intensity $I_{\rm i}^\alpha (x_s)$ as a function of scaled position in embryo $\alpha$.  In each experiment, we normalize by assuming that the minimum mean concentration is zero and we choose units such that the maximum mean concentration is one.  This defines
\begin{equation}
    g_{\rm i}^\alpha (x_s) = {1\over{S_{\rm i}}}\left[ I_{\rm i}^\alpha (x_s) - B_{\rm i}\right],
\end{equation}
where the background is
\begin{equation}
    B_{\rm i} = \min_{x_s} \langle   I_{\rm i}^\alpha (x_s) \rangle
\end{equation}
and the scale factor is 
\begin{equation}
    S_{\rm i} = \max_{x_s} \langle   I_{\rm i}^\alpha (x_s) \rangle
    - \min_{x_s}  \langle   I_{\rm i}^\alpha (x_s) \rangle .
\end{equation}
Importantly there is no freedom to normalize the profiles measured in individual embryos, which would distort our estimates of noise and variability \cite{gregor+al_07b}.

Data on three pair-rule genes---Eve, Prd, and Run---are taken from Ref. \cite{petkova+al_19}.  We fit the sum of seven Gaussians to each profile, identifying stripe positions with the centers of the Gaussians.  We have also used the average peak profiles as templates \cite{antonetti+al_18} and made more restricted fits to small segments of the peaks \cite{mcgough+al_23}; results are the same with all three methods.  Corrections for the drift of peak position vs.~time in nc14 were made as described in the main text.

\begin{figure}
\includegraphics[width=\linewidth]{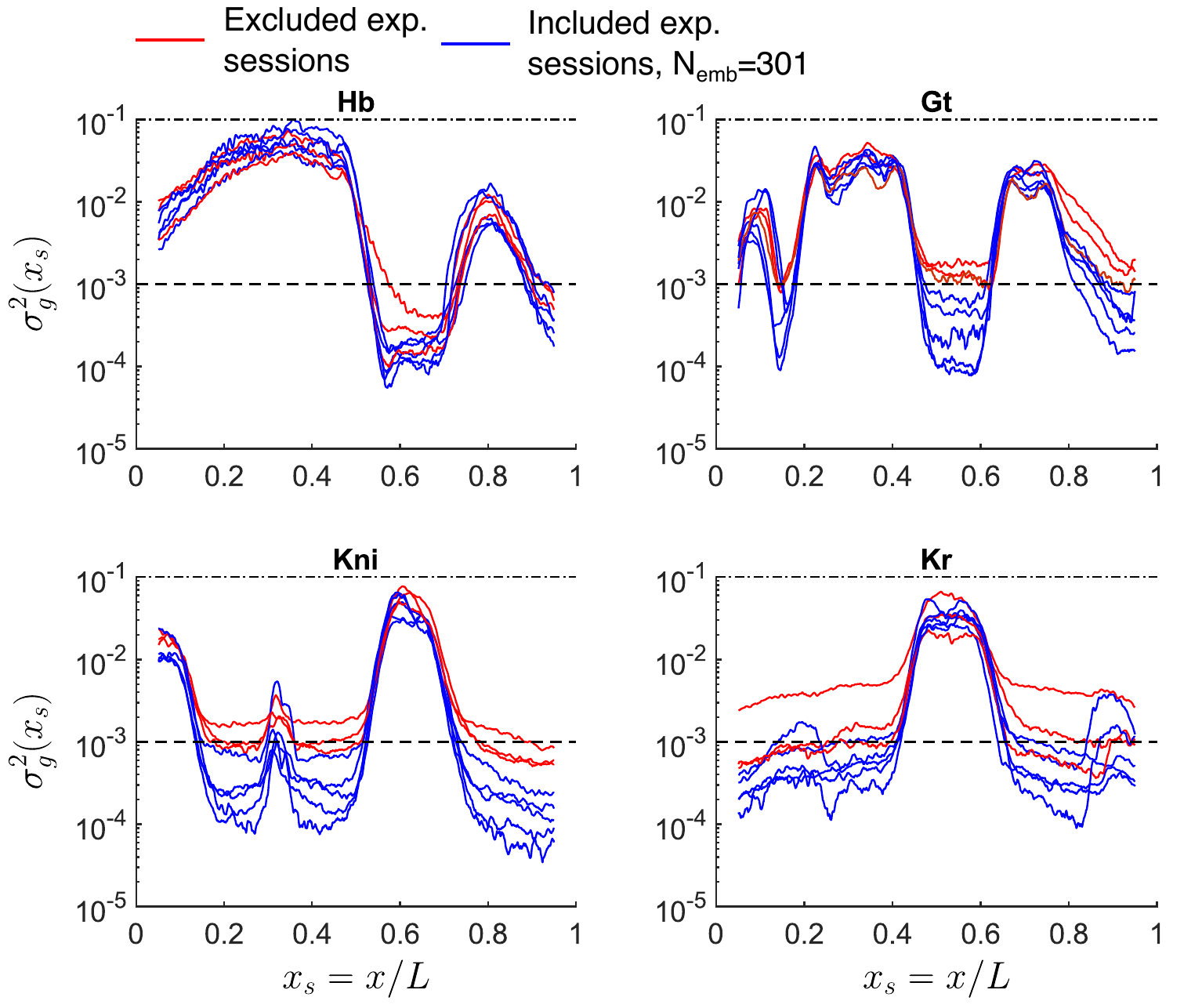}
\caption{{\bf Apparent variance of gap gene expression across multiple experimental sessions.}  Results from eight sessions are largely reproducible for each of the four genes.  In regions where mean expression levels are near zero, variances typically are $\sigma_g^2 \ll 10^{-3}$, except in a handful of sessions with highly variable backgrounds; these are excluded from further analysis.  \label{fig:backgrounds}}
\end{figure}

Simultaneous measurements on all four gap genes also were drawn from experiments described in Ref.~\cite{petkova+al_19}.  Because the analyses done here are so demanding of data, we tried to merge data from as many independent experimental sessions as possible.  Most quantities are extremely reproducible from session to session, but in a handful of sessions, we found anomalously large variations in background fluorescence across the individual embryos.  Concretely, if we measure the variance of expression levels for the individual genes, we typically find that $\sigma_g^2(x_s) \ll 10^{-3}$ in regions of $x_s$ with near zero mean (Fig.~\ref{fig:backgrounds}).  In a few sessions, these fluctuations in background are much larger, and these sessions are excluded; more precisely, since all genes are measured simultaneously, excess background variance in one gene is sufficient to exclude those data.  This leaves five independent sessions with a total of $N_{\rm em} = 301$ embryos which we pool into one data set for all further analyses.

For the analysis of gap gene expression boundaries, we mark the points that are half-maximal along the sharp slopes, as indicated in Fig.~\ref{fig:2}.  For the weak peak of Kni expression near $x_s = 0.33$ we fitted a Gaussian profile and took the marker as the center of the Gaussian.

Gap gene profiles vary slowly but significantly throughout nc14.  If we don't treat this variation explicitly it can be confused for noise, resulting in a substantial overestimate of the variances and entropies.  To separate the temporal dynamics of the gap genes from their noise level, we follow Ref.~\cite{dubuis+al_13a} and detrend the variations at each position, generalizing the treatment of the stripe positions in Fig.~\ref{fig:1}D. The alternative is to focus only on a small window of time \cite{petkova+al_19}, but this limits the size of the data set we can use.  

Another systematic source of variation is the dependence of gap gene profiles on the dorsal-ventral coordinate \cite{dubuis+al_13a}.  Previous work thus has been very strict in analyzing embryos with narrowly defined orientations.  To expand our data set we are less strict, but this is problematic for the Kni profiles in the range $0.15 < x_s < 0.45$, which contains a small peak. When analyzing Kni alone, or all four gap genes together, we exclude this region.  The alternative is to analyze the other three genes together across the full length of the anterior--posterior axis; results for $\Delta I$ are the same.

\begin{figure}
\includegraphics[width=\linewidth]{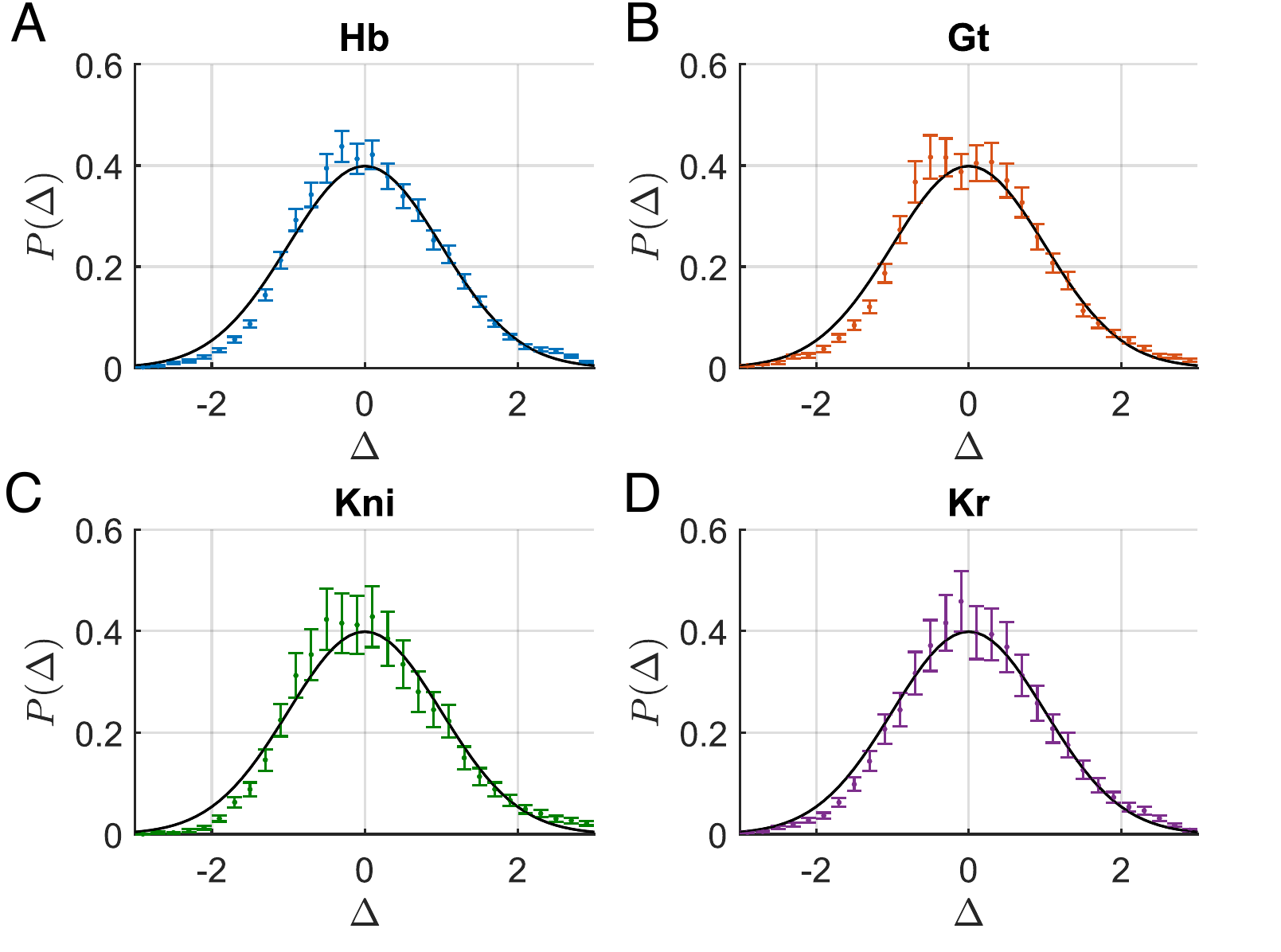}
\caption{{\bf Fluctuations in gap gene expression are approximately Gaussian.} 
Distributions of z-scored fluctuations, as in Eq.~(\ref{g-z}), are estimated for each individual gap gene, pooled across positions and embryos; error bars are standard deviations.  Black curves are Gaussians with zero mean and unit variance. \label{fig:gauss}}
\end{figure}

An important assumption for our analysis is that the distribution of gene expression at a given anterior--posterior position is Gaussian, as shown previously \cite{dubuis+al_13b}. For completeness, we reproduce this result for our larger data set. In a single embryo $\alpha$ we observe a gene expression level $g^\alpha(x_s)$ at scaled position $x_s$.  We  compute the mean and standard deviation across all the embryos in our ensemble and  define normalized deviations or z-scores
\begin{equation}
\Delta^\alpha(x_s) = {{g^\alpha(x_s) - \langle g(x_s)\rangle}\over{\sigma_g (x_s)}}.
\label{g-z}
\end{equation}
We pool across all $\alpha = 1,\, 2,\, \cdots ,\, N_{\rm em}$ embryos and across all positions $x_s$ to estimate the probability density $P(\Delta )$.  Results are in Fig.~\ref{fig:gauss} for each of the four gap genes.  

Finally, measurements of Bicoid concentration are taken from fluorescent imaging of live embryos expressions a Bicoid-GFP fusion  \cite{liu+al_13}; we consider only strain 2XA, which has the Bcd dosage closest to that of wild-type flies. 
With live measurements we can choose a time window, in this case $t=16\pm2\, \rm{min}$ after the start of nc14, avoiding any temporal detrending while still including $N_{\rm em} = 582$ embryos.  Some measurements with missing data points along the length of the embryo were excluded from this set.

\section{Estimating $\Delta I$ from limited data}
\label{app:finitedata}

Entropy and information depend on probability distributions, not just on their moments, and thus are especially difficult to estimate from a finite sample of data.  Further, the entropy is a nonlinear function of the probabilities, and so random errors in probability become systematic errors in our estimates of information.  This problem was appreciated in the very first attempts to use information theory in the analysis of experiments on biological systems \cite{miller_55}.  In the subsequent decades, many approaches have been developed, driven especially by the analysis of neural coding. The approach we take here follows the discussion in Appendix A.8 of Ref.~\cite{bialek_12}.  Rather than just saying that we follow established methods, we repeat some details that can be found in other contexts in hopes that our presentation thus will be more accessible.

If we estimate an information-theoretic quantity such as $\Delta I$ in Eq.~(\ref{deltaIdef2}) based on data from measurements in $N_{\rm em}$ independent embryos, then with any simple estimation procedure our estimates will be biased:
\begin{equation}
    \Delta I = \Delta I_{\infty} + {{A(N_{\rm bins})}\over {N_{\rm em}}} + {{B(N_{\rm bins})}\over {N_{\rm em}^2}} + \cdots .
    \label{IvsNem}
\end{equation}
Here $\Delta I_{\infty}$ is the true value of $\Delta I$ which we would observe if we could collect an infinite number of samples.  The notation reminds us that if we make bins along some continuous axis, then the size of the corrections at finite $N_{\rm em}$ depend on the number of bins $N_{\rm bins}$.  With more bins the corrections are larger, which means that a naive estimate with a fixed number of embryos will depend on the bin size.  The hope is that we can find a regime in which the extrapolated $\Delta I_\infty$ is independent of $N_{\rm bins}$.

It is important that Eq.~(\ref{IvsNem}) is not just a guess, but a prediction that can be derived theoretically.  Theory also gives values for the coefficients $A$ and $B$, but these depend on details such as the independence of samples; the form is more general.  This suggests a strategy in which we vary the number of embryos that we include in our analysis and look for the predicted systematic dependence on $N_{\rm em}$.  If we can see this behavior then we can feel confident in fitting to Eq.~(\ref{IvsNem}) and extracting an estimate $\Delta I_\infty$ \cite{bialek_12}. 

\begin{figure}
\includegraphics[width=\linewidth]{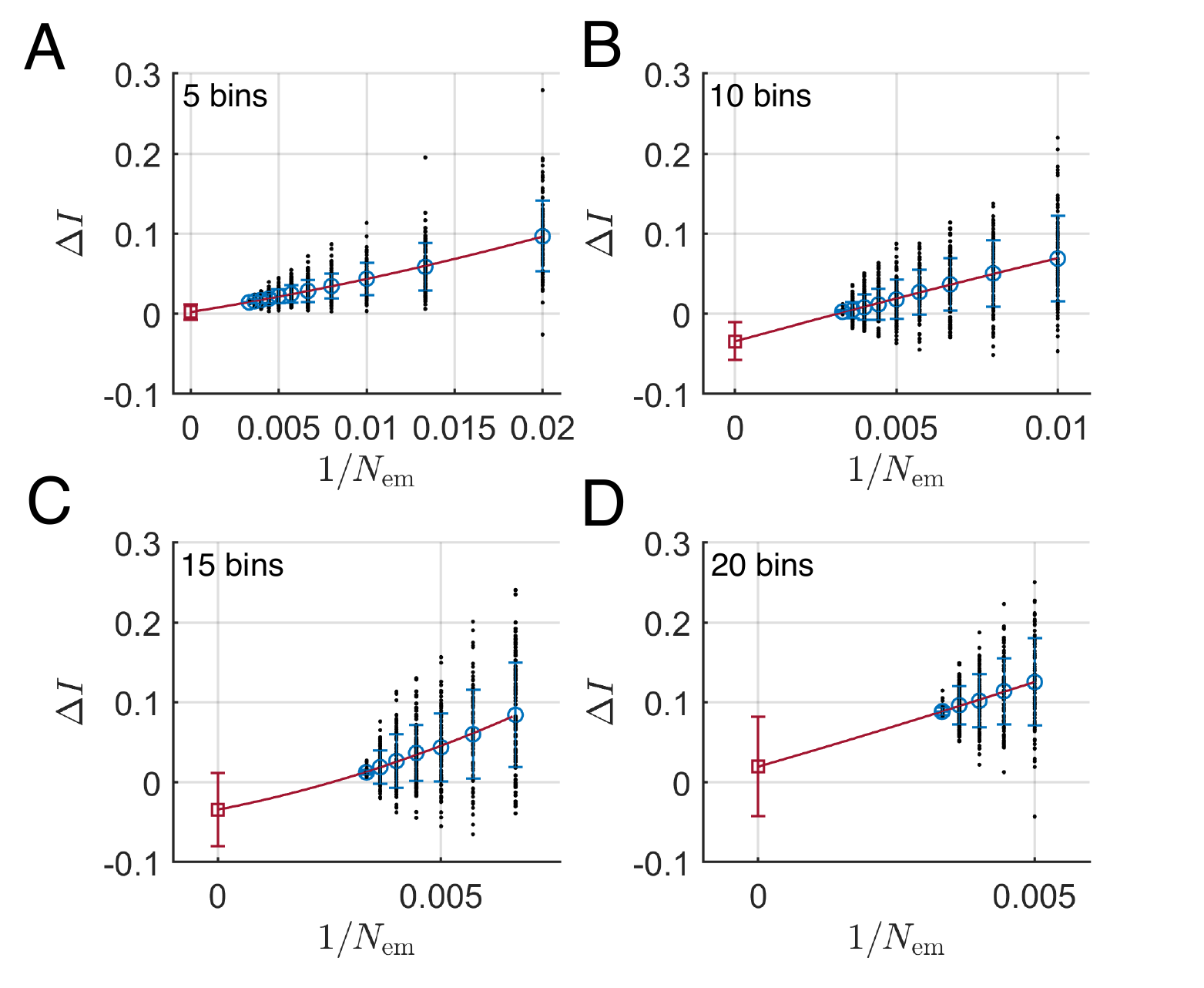}
\caption{{\bf Consistent estimates of $\Delta I$ with varying $N_{\rm bins}$.} (A) Repeats the results of Fig \ref{fig:Igap}A on $\Delta I$ vs.~$N_{\rm em}$ for Hb, analyzed with $N_{\rm bins} = 5$ adaptive bins along the $L$ axis. (B) As in (A) with $N_{\rm bins} = 10$; (C) with $N_{\rm bins} = 15$; and (D) with $N_{\rm bins} = 20$.  Systematic errors are large at fixed $N_{\rm em}$ and increasing $N_{\rm bins}$, but the extrapolation $N_{\rm em}\rightarrow \infty$ is within error bars of $\Delta I = 0$ in each case.  \label{fig:Lbins}}
\end{figure}

This estimation procedure is illustrated by Fig.~\ref{fig:Igap} in the main text and by Fig.~\ref{fig:Lbins}.  When we vary the number of embryos that we include in our analysis, we can choose at random from the total number available, so we have a path also to estimating error bars (below).  In Fig.~\ref{fig:Igap}A we analyze $\Delta I$ for the spatial profiles of Hb expression using the Gaussian approximation of Eq.~(\ref{DeltaI_vars}).  We have to make estimates of the variance as a function of the scaled coordinate $x_s$ and the embryo length $L$.  As explained in the main text, we choose bins of $\Delta x_s = 0.01$, consistent with the observed precision of the pair-rule stripes and with earlier work \cite{dubuis+al_13b,petkova+al_19}.  Along the $L$ axis we use adaptive bins, that is bins with boundaries chosen so that the number of embryos in each bin is as nearly equal as possible; these bins are chosen based on the full experimental ensembles, and not readjusted as we choose smaller samples at random.

In Figure \ref{fig:Igap}A we have chosen $N_{\rm bins} =5$ adaptive bins along the $L$ axis, and we see a clean depending of $\Delta I$ on $N_{\rm em}$ as predicted in Eq.~(\ref{IvsNem}).  The dependence on $N_{\rm em}$ is dominated by the linear term $A$, although some curvature is detectable.  The quality of the fit to Eq.~(\ref{IvsNem}) is very good, and the extrapolation is to $\Delta I_\infty =0$ with a precision of better than $0.01\,{\rm bits}$.  

In Figure \ref{fig:Lbins} we see how the plot of $\Delta I$ vs.~$N_{\rm em}$  changes as we vary $N_{\rm bins} = 5,\, 10,\, 15,\, 20$.  With more bins, there are fewer embryos in each bin, which means that the minimum number of embryos that we need for a meaningful analysis is larger.  Increasing the number of bins might reveal otherwise hidden information, but also increases the size of systematic errors.  Comparing across the panels in Fig.~\ref{fig:Lbins} we see that at fixed $N_{\rm em}$ the apparent $\Delta I$ increases with $N_{\rm bins}$, and if we didn't explore the full dependence on $N_{\rm em}$ we might be tempted to conclude that we are indeed revealing extra information.  But this is not the case, since the plots at all values of $N_{\rm bins}$ extrapolate to zero within error bars.

\begin{figure}[h]
\centering{\includegraphics[width=\linewidth]{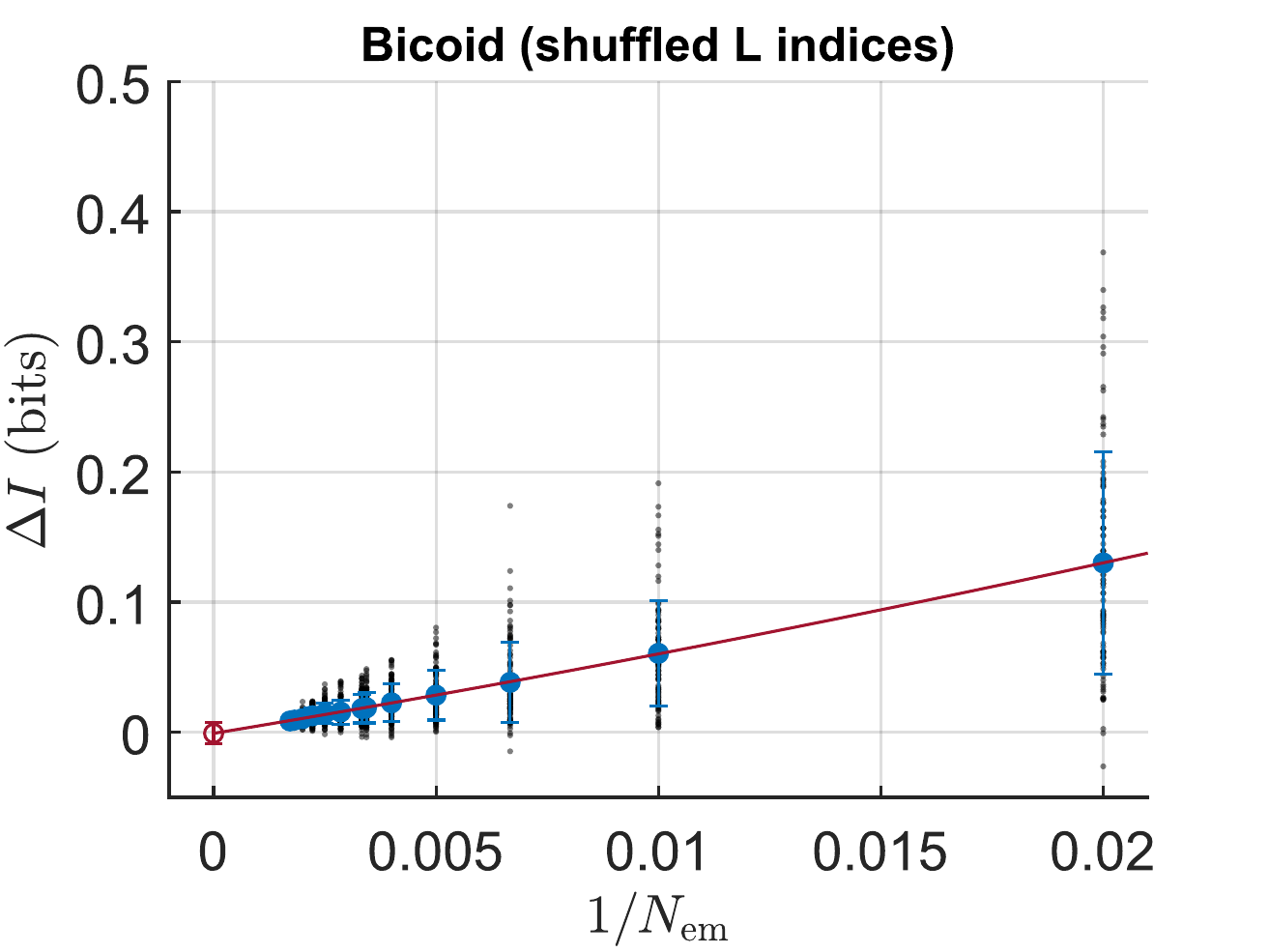}}
\caption{{\bf Recovering $\Delta I =0$ in shuffled data.} We permute the lengths $L$ of the embryos at random and repeat the analysis of the Bcd profiles.  While the real data extrapolate to nonzero $\Delta I$ (Fig \ref{fig:3}C), here we recover $\Delta I = 0$ as expected. \label{fig:extrap_to_zero}}
\end{figure}

One useful test of these extrapolation methods is to be sure that we arrive at zero information in those cases where we know that the answer must be zero.  As an example, if we randomly permute or shuffle the data we can break correlations that lead to nonzero information.  In this case, if we randomly reassign lengths $L$ to the embryos, then we must have $\Delta I = 0$.  This is illustrated in Fig.~\ref{fig:extrap_to_zero}, using the example of Bcd.  Here the real data extrapolate to a nonzero value of $\Delta I$ (Fig.~\ref{fig:3}), and when we shuffle we still see significantly nonzero values at $N_{\rm em} \sim 100$.  But using the whole $N_{\rm em}$ dependence we can see this extrapolates smoothly to zero, as it should.

\begin{figure}[b]
\includegraphics[width=\linewidth]{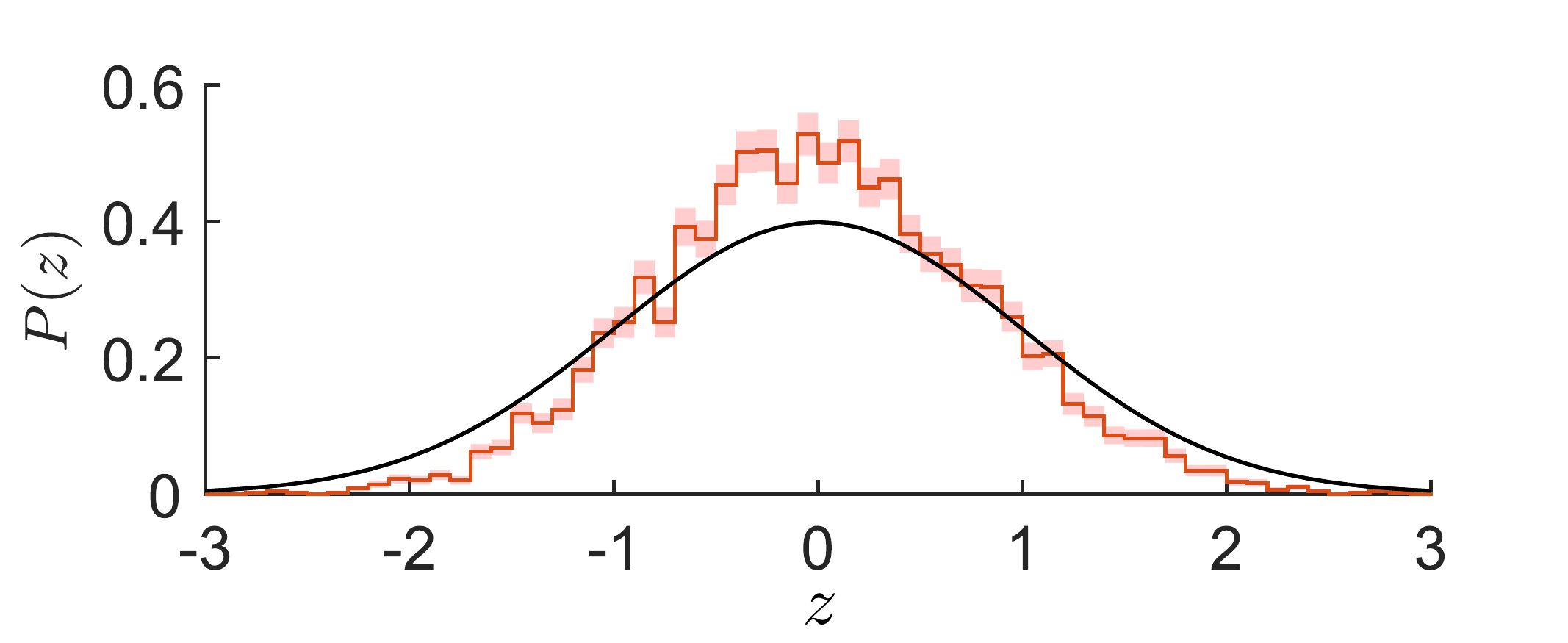}
\caption{{\bf Distribution of errors in $\Delta I$.} The distribution of errors of $\Delta I$ was estimated by repeating 5000 times the entire procedure leading to Fig.~\ref{fig:Hb}A, on shuffled versions of the Hb data, where we know the true value of $\Delta I$ is 0 bits.  We calculate the z-score based on our estimates of $\Delta I$ and error bars $\sigma_{\Delta I}$. Red histogram shows the probability distribution $P(z)$ in bins of size $\Delta z  = 0.1$. Shaded area is the uncertainty estimated through bootstrapping; black line is a Gaussian distribution with $\langle z \rangle = 0$ and $\langle z^2\rangle = 1$.   \label{fig:InfErr}}
\end{figure}

An essential part of this analysis is the estimation of error bars.  For reasonably large $N_{\rm em}$ the systematic and random errors are additive, and the variance of random errors scales  $\propto 1/N_{\rm em}$ as usual.  This means that if we compute the variance of $\Delta I$ across random halves of the data, and divide by two, we should have a good estimate of the variance in $\Delta I$ based on all of the data. If this error bar $\sigma_{\Delta I}$ is correct, and our extrapolation is consistent, then when the true $\Delta I$ is zero, as with shuffled data, we can form a z-score, $z = \Delta_\infty/\sigma_{\Delta I}$, and $z$ should be Gaussian with zero mean and unit variance. We can test this because the extrapolation procedure involves taking random subsamples of the data, each of which generates a slightly different value of $\Delta I_\infty$.  Figure \ref{fig:InfErr} shows the distribution $P(z)$ obtained from a shuffled version of the Hb data, illustrating a good match to the expected Gaussian distribution; the deviation is a bias toward smaller $z$, suggesting that our estimates of the error bars may be a bit conservative.

\begin{figure}
\centering{\includegraphics[width=\linewidth]{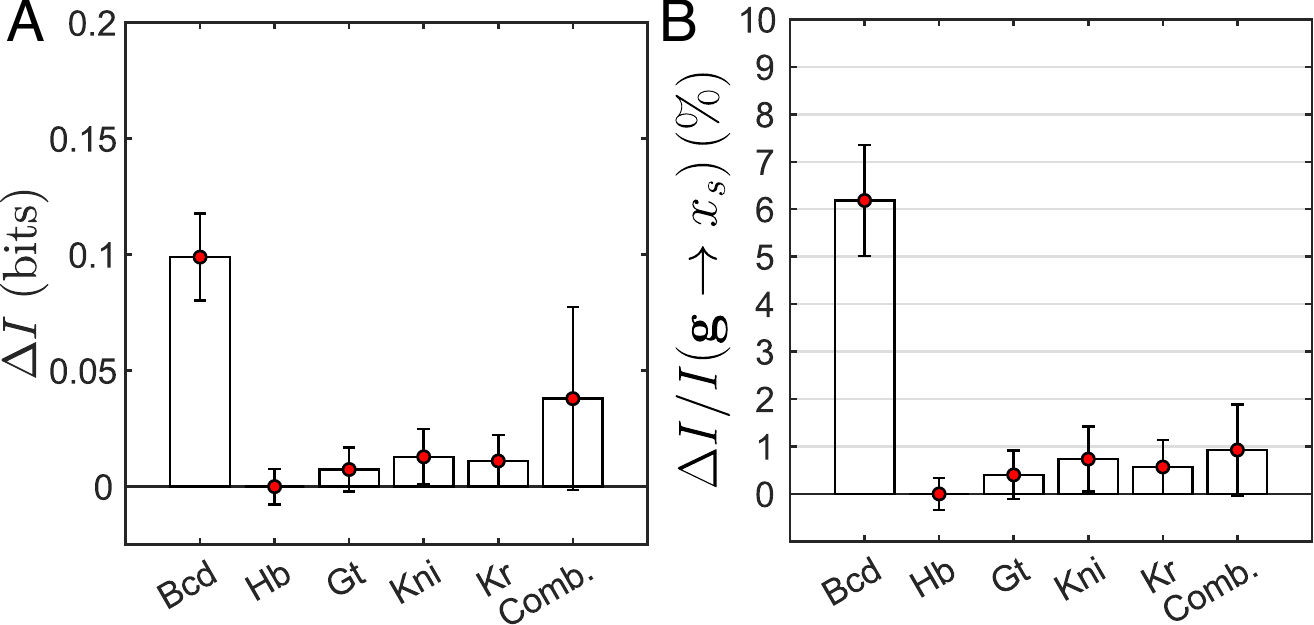}}
\caption{{\bf Summary of deviations from scale invariance.} (A) Extrapolated $\Delta I$ with error bars for Bicoid, each gap gene individually, and the combination of all gap genes. (B) Deviation from scale invariance as a percentage of the information about scaled position  \cite{dubuis+al_13b,tkacik+al_15}. \label{fig:DeltaIsumm}}
\end{figure}

Now that we have control over both the systematic and random errors in estimating $\Delta I$, we summarize the results.  We have done separate analyses for each of the gap genes, for all four gap genes together, and for the maternal input Bicoid.  As we see in Fig.~\ref{fig:DeltaIsumm}A, all results for the gap genes are consistent with $\Delta I = 0 \,{\rm bits}$ within error bars, while for Bicoid we find a significantly nonzero value.  These quantities are perhaps best expressed as fractions of the information conveyed about scaled position, as shown in Fig.~\ref{fig:DeltaIsumm}B; estimates of $I({\mathbf g} \rightarrow x_s)$ are from Refs.~\cite{dubuis+al_13b,tkacik+al_15}.

\vfill\newpage

\bibliography{embryo_bibliography}

%apsrev4-2.bst 2019-01-14 (MD) hand-edited version of apsrev4-1.bst
%Control: key (0)
%Control: author (8) initials jnrlst
%Control: editor formatted (1) identically to author
%Control: production of article title (0) allowed
%Control: page (0) single
%Control: year (1) truncated
%Control: production of eprint (0) enabled
\begin{thebibliography}{65}%
\makeatletter
\providecommand \@ifxundefined [1]{%
 \@ifx{#1\undefined}
}%
\providecommand \@ifnum [1]{%
 \ifnum #1\expandafter \@firstoftwo
 \else \expandafter \@secondoftwo
 \fi
}%
\providecommand \@ifx [1]{%
 \ifx #1\expandafter \@firstoftwo
 \else \expandafter \@secondoftwo
 \fi
}%
\providecommand \natexlab [1]{#1}%
\providecommand \enquote  [1]{``#1''}%
\providecommand \bibnamefont  [1]{#1}%
\providecommand \bibfnamefont [1]{#1}%
\providecommand \citenamefont [1]{#1}%
\providecommand \href@noop [0]{\@secondoftwo}%
\providecommand \href [0]{\begingroup \@sanitize@url \@href}%
\providecommand \@href[1]{\@@startlink{#1}\@@href}%
\providecommand \@@href[1]{\endgroup#1\@@endlink}%
\providecommand \@sanitize@url [0]{\catcode `\\12\catcode `\$12\catcode
  `\&12\catcode `\#12\catcode `\^12\catcode `\_12\catcode `\%12\relax}%
\providecommand \@@startlink[1]{}%
\providecommand \@@endlink[0]{}%
\providecommand \url  [0]{\begingroup\@sanitize@url \@url }%
\providecommand \@url [1]{\endgroup\@href {#1}{\urlprefix }}%
\providecommand \urlprefix  [0]{URL }%
\providecommand \Eprint [0]{\href }%
\providecommand \doibase [0]{https://doi.org/}%
\providecommand \selectlanguage [0]{\@gobble}%
\providecommand \bibinfo  [0]{\@secondoftwo}%
\providecommand \bibfield  [0]{\@secondoftwo}%
\providecommand \translation [1]{[#1]}%
\providecommand \BibitemOpen [0]{}%
\providecommand \bibitemStop [0]{}%
\providecommand \bibitemNoStop [0]{.\EOS\space}%
\providecommand \EOS [0]{\spacefactor3000\relax}%
\providecommand \BibitemShut  [1]{\csname bibitem#1\endcsname}%
\let\auto@bib@innerbib\@empty
%</preamble>
\bibitem [{\citenamefont {Ishimatsu}\ \emph {et~al.}(2018)\citenamefont
  {Ishimatsu}, \citenamefont {Hiscock}, \citenamefont {M.}, \citenamefont
  {Sari}, \citenamefont {Lischer}, \citenamefont {Richmond}, \citenamefont
  {Bessho}, \citenamefont {Matsui},\ and\ \citenamefont
  {Megason}}]{ishimatsu+al_18}%
  \BibitemOpen
  \bibfield  {author} {\bibinfo {author} {\bibfnamefont {K.}~\bibnamefont
  {Ishimatsu}}, \bibinfo {author} {\bibfnamefont {T.~W.}\ \bibnamefont
  {Hiscock}}, \bibinfo {author} {\bibfnamefont {C.~Z.}\ \bibnamefont {M.}},
  \bibinfo {author} {\bibfnamefont {D.~W.~K.}\ \bibnamefont {Sari}}, \bibinfo
  {author} {\bibfnamefont {K.}~\bibnamefont {Lischer}}, \bibinfo {author}
  {\bibfnamefont {D.~L.}\ \bibnamefont {Richmond}}, \bibinfo {author}
  {\bibfnamefont {Y.}~\bibnamefont {Bessho}}, \bibinfo {author} {\bibfnamefont
  {T.}~\bibnamefont {Matsui}},\ and\ \bibinfo {author} {\bibfnamefont {S.~G.}\
  \bibnamefont {Megason}},\ }\bibfield  {title} {\bibinfo {title} {Size-reduced
  embryos reveal a gradient scaling based mechanism for zebrafish somite
  formation},\ }\href@noop {} {\bibfield  {journal} {\bibinfo  {journal}
  {Development}\ }\textbf {\bibinfo {volume} {145}},\ \bibinfo {pages}
  {dev161257} (\bibinfo {year} {2018})}\BibitemShut {NoStop}%
\bibitem [{\citenamefont {Almuedo-Castillo}\ \emph {et~al.}(2018)\citenamefont
  {Almuedo-Castillo}, \citenamefont {Bl\"a{\ss}le}, \citenamefont {M\"orsdorf},
  \citenamefont {Marcon}, \citenamefont {Soh}, \citenamefont {Rogers},
  \citenamefont {Schier},\ and\ \citenamefont
  {M\"uller}}]{almuedo-castillo+al_18}%
  \BibitemOpen
  \bibfield  {author} {\bibinfo {author} {\bibfnamefont {M.}~\bibnamefont
  {Almuedo-Castillo}}, \bibinfo {author} {\bibfnamefont {A.}~\bibnamefont
  {Bl\"a{\ss}le}}, \bibinfo {author} {\bibfnamefont {D.}~\bibnamefont
  {M\"orsdorf}}, \bibinfo {author} {\bibfnamefont {L.}~\bibnamefont {Marcon}},
  \bibinfo {author} {\bibfnamefont {G.~H.}\ \bibnamefont {Soh}}, \bibinfo
  {author} {\bibfnamefont {K.~W.}\ \bibnamefont {Rogers}}, \bibinfo {author}
  {\bibfnamefont {A.~F.}\ \bibnamefont {Schier}},\ and\ \bibinfo {author}
  {\bibfnamefont {P.}~\bibnamefont {M\"uller}},\ }\href@noop {} {\bibfield
  {journal} {\bibinfo  {journal} {Nature Cell Biology}\ }\textbf {\bibinfo
  {volume} {20}},\ \bibinfo {pages} {1032} (\bibinfo {year}
  {2018})}\BibitemShut {NoStop}%
\bibitem [{\citenamefont {Leibovich}\ \emph {et~al.}(2020)\citenamefont
  {Leibovich}, \citenamefont {Edri}, \citenamefont {Klein}, \citenamefont
  {Moody},\ and\ \citenamefont {Fainsod}}]{leibovich+al_20}%
  \BibitemOpen
  \bibfield  {author} {\bibinfo {author} {\bibfnamefont {A.}~\bibnamefont
  {Leibovich}}, \bibinfo {author} {\bibfnamefont {T.}~\bibnamefont {Edri}},
  \bibinfo {author} {\bibfnamefont {S.~L.}\ \bibnamefont {Klein}}, \bibinfo
  {author} {\bibfnamefont {S.~A.}\ \bibnamefont {Moody}},\ and\ \bibinfo
  {author} {\bibfnamefont {A.}~\bibnamefont {Fainsod}},\ }\bibfield  {title}
  {\bibinfo {title} {Natural size variation among embryos leads to the
  corresponding scaling in gene expression},\ }\href@noop {} {\bibfield
  {journal} {\bibinfo  {journal} {Developmental Biology}\ }\textbf {\bibinfo
  {volume} {462}},\ \bibinfo {pages} {165} (\bibinfo {year}
  {2020})}\BibitemShut {NoStop}%
\bibitem [{\citenamefont {Huxley}(1932)}]{huxley_32}%
  \BibitemOpen
  \bibfield  {author} {\bibinfo {author} {\bibfnamefont {J.~S.}\ \bibnamefont
  {Huxley}},\ }\href@noop {} {\emph {\bibinfo {title} {Problems of Relative
  Growth}}}\ (\bibinfo  {publisher} {Meuthen},\ \bibinfo {address} {London},\
  \bibinfo {year} {1932})\BibitemShut {NoStop}%
\bibitem [{\citenamefont {McMahon}\ and\ \citenamefont
  {Bonner}(1983)}]{mcmahon+bonner_1983}%
  \BibitemOpen
  \bibfield  {author} {\bibinfo {author} {\bibfnamefont {T.~A.}\ \bibnamefont
  {McMahon}}\ and\ \bibinfo {author} {\bibfnamefont {J.~T.}\ \bibnamefont
  {Bonner}},\ }\href@noop {} {\emph {\bibinfo {title} {On Size and Life}}}\
  (\bibinfo  {publisher} {Scientific American Library},\ \bibinfo {address}
  {New York},\ \bibinfo {year} {1983})\BibitemShut {NoStop}%
\bibitem [{\citenamefont {West}\ \emph {et~al.}(1997)\citenamefont {West},
  \citenamefont {Brown},\ and\ \citenamefont {Enquist}}]{west+al_97}%
  \BibitemOpen
  \bibfield  {author} {\bibinfo {author} {\bibfnamefont {G.~B.}\ \bibnamefont
  {West}}, \bibinfo {author} {\bibfnamefont {J.~H.}\ \bibnamefont {Brown}},\
  and\ \bibinfo {author} {\bibfnamefont {B.~J.}\ \bibnamefont {Enquist}},\
  }\bibfield  {title} {\bibinfo {title} {A general model for the origin of
  allometric scaling laws in biology},\ }\href@noop {} {\bibfield  {journal}
  {\bibinfo  {journal} {Science}\ }\textbf {\bibinfo {volume} {276}},\ \bibinfo
  {pages} {122} (\bibinfo {year} {1997})}\BibitemShut {NoStop}%
\bibitem [{\citenamefont {Kepler}(2010)}]{kepler}%
  \BibitemOpen
  \bibfield  {author} {\bibinfo {author} {\bibfnamefont {J.}~\bibnamefont
  {Kepler}},\ }\href@noop {} {\emph {\bibinfo {title} {The Six--Cornered
  Snowflake: A New Year's Gift}}}\ (\bibinfo  {publisher} {Paul Dry Books},\
  \bibinfo {address} {Philadelphia, PA},\ \bibinfo {year} {2010})\ \bibinfo
  {note} {translated from the 1611 Edition by J Bromberg}\BibitemShut {NoStop}%
\bibitem [{\citenamefont {B\'enard}(1900)}]{benard_00}%
  \BibitemOpen
  \bibfield  {author} {\bibinfo {author} {\bibfnamefont {H.}~\bibnamefont
  {B\'enard}},\ }\bibfield  {title} {\bibinfo {title} {Les tourbillions
  cellulaires dans une nappe liquide transportent de la chaleur par convection
  en regime permanent},\ }\href@noop {} {\bibfield  {journal} {\bibinfo
  {journal} {Ann Chim Phys}\ }\textbf {\bibinfo {volume} {7}},\ \bibinfo
  {pages} {62} (\bibinfo {year} {1900})}\BibitemShut {NoStop}%
\bibitem [{\citenamefont {{Lord Rayleigh}}(1916)}]{rayleigh_16}%
  \BibitemOpen
  \bibfield  {author} {\bibinfo {author} {\bibnamefont {{Lord Rayleigh}}},\
  }\bibfield  {title} {\bibinfo {title} {On convection currents in a horizontal
  layer of fluid, when the higher temperature is on the under side},\
  }\href@noop {} {\bibfield  {journal} {\bibinfo  {journal} {Phil Mag}\
  }\textbf {\bibinfo {volume} {32}},\ \bibinfo {pages} {529} (\bibinfo {year}
  {1916})}\BibitemShut {NoStop}%
\bibitem [{\citenamefont {Flesselles}\ \emph {et~al.}(1991)\citenamefont
  {Flesselles}, \citenamefont {Simon},\ and\ \citenamefont
  {Libchaber}}]{flesselles+al_91}%
  \BibitemOpen
  \bibfield  {author} {\bibinfo {author} {\bibfnamefont {J.-M.}\ \bibnamefont
  {Flesselles}}, \bibinfo {author} {\bibfnamefont {A.~J.}\ \bibnamefont
  {Simon}},\ and\ \bibinfo {author} {\bibfnamefont {A.~J.}\ \bibnamefont
  {Libchaber}},\ }\bibfield  {title} {\bibinfo {title} {Dynamics of
  one-dimensional interfaces: An experimentalist's view},\ }\href@noop {}
  {\bibfield  {journal} {\bibinfo  {journal} {Advances in Physics}\ }\textbf
  {\bibinfo {volume} {40}},\ \bibinfo {pages} {1} (\bibinfo {year}
  {1991})}\BibitemShut {NoStop}%
\bibitem [{\citenamefont {Cross}\ and\ \citenamefont
  {Hohenberg}(1993)}]{cross+hohenberg_93}%
  \BibitemOpen
  \bibfield  {author} {\bibinfo {author} {\bibfnamefont {M.~C.}\ \bibnamefont
  {Cross}}\ and\ \bibinfo {author} {\bibfnamefont {P.~C.}\ \bibnamefont
  {Hohenberg}},\ }\bibfield  {title} {\bibinfo {title} {Pattern formation
  outside of equilibrium},\ }\href@noop {} {\bibfield  {journal} {\bibinfo
  {journal} {Reviews of Modern Physics}\ }\textbf {\bibinfo {volume} {65}},\
  \bibinfo {pages} {851} (\bibinfo {year} {1993})}\BibitemShut {NoStop}%
\bibitem [{\citenamefont {Lappa}(2009)}]{lappa_09}%
  \BibitemOpen
  \bibfield  {author} {\bibinfo {author} {\bibfnamefont {M.}~\bibnamefont
  {Lappa}},\ }\href@noop {} {\emph {\bibinfo {title} {Thermal Convection:
  Patterns, Evolution and Stability}}}\ (\bibinfo  {publisher} {John Wiley and
  Sons},\ \bibinfo {address} {New York},\ \bibinfo {year} {2009})\BibitemShut
  {NoStop}%
\bibitem [{\citenamefont {Langer}(1989)}]{langer_89}%
  \BibitemOpen
  \bibfield  {author} {\bibinfo {author} {\bibfnamefont {J.~S.}\ \bibnamefont
  {Langer}},\ }\bibfield  {title} {\bibinfo {title} {Dendrites, viscous
  fingers, and the theory of pattern formation},\ }\href@noop {} {\bibfield
  {journal} {\bibinfo  {journal} {Science}\ }\textbf {\bibinfo {volume}
  {243}},\ \bibinfo {pages} {1150} (\bibinfo {year} {1989})}\BibitemShut
  {NoStop}%
\bibitem [{\citenamefont {Turing}(1952)}]{turing_52}%
  \BibitemOpen
  \bibfield  {author} {\bibinfo {author} {\bibfnamefont {A.~M.}\ \bibnamefont
  {Turing}},\ }\bibfield  {title} {\bibinfo {title} {The chemical basis of
  morphogenesis},\ }\href@noop {} {\bibfield  {journal} {\bibinfo  {journal}
  {Philos Trans R Soc Lond B}\ }\textbf {\bibinfo {volume} {237}},\ \bibinfo
  {pages} {37} (\bibinfo {year} {1952})}\BibitemShut {NoStop}%
\bibitem [{\citenamefont {Scott}\ and\ \citenamefont
  {Carroll}(1987)}]{scott+carroll_87}%
  \BibitemOpen
  \bibfield  {author} {\bibinfo {author} {\bibfnamefont {M.~P.}\ \bibnamefont
  {Scott}}\ and\ \bibinfo {author} {\bibfnamefont {S.~B.}\ \bibnamefont
  {Carroll}},\ }\bibfield  {title} {\bibinfo {title} {{The segmentation and
  homeotic gene network in early {\em Drosophila} development}},\ }\href@noop
  {} {\bibfield  {journal} {\bibinfo  {journal} {Cell}\ }\textbf {\bibinfo
  {volume} {51}},\ \bibinfo {pages} {P689} (\bibinfo {year}
  {1987})}\BibitemShut {NoStop}%
\bibitem [{\citenamefont {Jaeger}\ and\ \citenamefont
  {Verd}(2020)}]{jaeger+verd_20}%
  \BibitemOpen
  \bibfield  {author} {\bibinfo {author} {\bibfnamefont {J.}~\bibnamefont
  {Jaeger}}\ and\ \bibinfo {author} {\bibfnamefont {B.}~\bibnamefont {Verd}},\
  }\bibfield  {title} {\bibinfo {title} {Dynamic positional information:
  Patterning mechanism versus precision in gradient-driven systems},\
  }\href@noop {} {\bibfield  {journal} {\bibinfo  {journal} {Current Topics in
  Developmental Biology}\ }\textbf {\bibinfo {volume} {137}},\ \bibinfo {pages}
  {219} (\bibinfo {year} {2020})}\BibitemShut {NoStop}%
\bibitem [{\citenamefont {Tka\v{c}ik}\ and\ \citenamefont
  {Gregor}(2021)}]{tkacik+gregor_21}%
  \BibitemOpen
  \bibfield  {author} {\bibinfo {author} {\bibfnamefont {G.}~\bibnamefont
  {Tka\v{c}ik}}\ and\ \bibinfo {author} {\bibfnamefont {T.}~\bibnamefont
  {Gregor}},\ }\bibfield  {title} {\bibinfo {title} {The many bits of
  positional information},\ }\href@noop {} {\bibfield  {journal} {\bibinfo
  {journal} {Development}\ }\textbf {\bibinfo {volume} {148}},\ \bibinfo
  {pages} {dev176065} (\bibinfo {year} {2021})}\BibitemShut {NoStop}%
\bibitem [{\citenamefont {Houchmandzadeh}\ \emph {et~al.}(2002)\citenamefont
  {Houchmandzadeh}, \citenamefont {Wieschaus},\ and\ \citenamefont
  {Leibler}}]{houchmandzadeh+al_02}%
  \BibitemOpen
  \bibfield  {author} {\bibinfo {author} {\bibfnamefont {B.}~\bibnamefont
  {Houchmandzadeh}}, \bibinfo {author} {\bibfnamefont {E.}~\bibnamefont
  {Wieschaus}},\ and\ \bibinfo {author} {\bibfnamefont {S.}~\bibnamefont
  {Leibler}},\ }\bibfield  {title} {\bibinfo {title} {{Establishment of
  developmental precision and proportions in the early {\em Drosophila}
  embryo}},\ }\href@noop {} {\bibfield  {journal} {\bibinfo  {journal}
  {Nature}\ }\textbf {\bibinfo {volume} {415}},\ \bibinfo {pages} {798}
  (\bibinfo {year} {2002})}\BibitemShut {NoStop}%
\bibitem [{\citenamefont {Surkova}\ \emph {et~al.}(2008)\citenamefont
  {Surkova}, \citenamefont {Kosman}, \citenamefont {Kozlov}, \citenamefont
  {Myasnikova}, \citenamefont {Samsonova}, \citenamefont {Spirov},
  \citenamefont {Vanario-Alonso}, \citenamefont {Samsonova}, \citenamefont
  {Reinitz} \emph {et~al.}}]{surkova+al_08}%
  \BibitemOpen
  \bibfield  {author} {\bibinfo {author} {\bibfnamefont {S.}~\bibnamefont
  {Surkova}}, \bibinfo {author} {\bibfnamefont {D.}~\bibnamefont {Kosman}},
  \bibinfo {author} {\bibfnamefont {K.}~\bibnamefont {Kozlov}}, \bibinfo
  {author} {\bibfnamefont {E.}~\bibnamefont {Myasnikova}}, \bibinfo {author}
  {\bibfnamefont {A.~A.}\ \bibnamefont {Samsonova}}, \bibinfo {author}
  {\bibfnamefont {A.}~\bibnamefont {Spirov}}, \bibinfo {author} {\bibfnamefont
  {C.~E.}\ \bibnamefont {Vanario-Alonso}}, \bibinfo {author} {\bibfnamefont
  {M.}~\bibnamefont {Samsonova}}, \bibinfo {author} {\bibfnamefont
  {J.}~\bibnamefont {Reinitz}}, \emph {et~al.},\ }\bibfield  {title} {\bibinfo
  {title} {{Characterization of the {\em Drosophila} segment determination
  morphome}},\ }\href@noop {} {\bibfield  {journal} {\bibinfo  {journal}
  {Developmental Biology}\ }\textbf {\bibinfo {volume} {313}},\ \bibinfo
  {pages} {844} (\bibinfo {year} {2008})}\BibitemShut {NoStop}%
\bibitem [{\citenamefont {Holloway}\ \emph {et~al.}(2006)\citenamefont
  {Holloway}, \citenamefont {Harrison}, \citenamefont {Kosman}, \citenamefont
  {Vanario-Alonso},\ and\ \citenamefont {Spirov}}]{holloway+al_06}%
  \BibitemOpen
  \bibfield  {author} {\bibinfo {author} {\bibfnamefont {D.~M.}\ \bibnamefont
  {Holloway}}, \bibinfo {author} {\bibfnamefont {L.~G.}\ \bibnamefont
  {Harrison}}, \bibinfo {author} {\bibfnamefont {D.}~\bibnamefont {Kosman}},
  \bibinfo {author} {\bibfnamefont {C.~E.}\ \bibnamefont {Vanario-Alonso}},\
  and\ \bibinfo {author} {\bibfnamefont {A.~V.}\ \bibnamefont {Spirov}},\
  }\bibfield  {title} {\bibinfo {title} {Analysis of pattern precision shows
  that drosophila segmentation develops substantial independence from gradients
  of maternal gene products},\ }\href@noop {} {\bibfield  {journal} {\bibinfo
  {journal} {Developmental Dynamics}\ }\textbf {\bibinfo {volume} {235}},\
  \bibinfo {pages} {2949} (\bibinfo {year} {2006})}\BibitemShut {NoStop}%
\bibitem [{\citenamefont {Lott}\ \emph {et~al.}(2007)\citenamefont {Lott},
  \citenamefont {Kreitman}, \citenamefont {Palsson}, \citenamefont
  {Alekseeva},\ and\ \citenamefont {Ludwig}}]{lott+al_07}%
  \BibitemOpen
  \bibfield  {author} {\bibinfo {author} {\bibfnamefont {S.~E.}\ \bibnamefont
  {Lott}}, \bibinfo {author} {\bibfnamefont {M.}~\bibnamefont {Kreitman}},
  \bibinfo {author} {\bibfnamefont {A.}~\bibnamefont {Palsson}}, \bibinfo
  {author} {\bibfnamefont {E.}~\bibnamefont {Alekseeva}},\ and\ \bibinfo
  {author} {\bibfnamefont {M.~Z.}\ \bibnamefont {Ludwig}},\ }\bibfield  {title}
  {\bibinfo {title} {{Canalization of segmentation and its evolution in {\em
  Drosophila}}},\ }\href@noop {} {\bibfield  {journal} {\bibinfo  {journal}
  {Proceedings of the National Academy of Sciences (USA)}\ }\textbf {\bibinfo
  {volume} {104}},\ \bibinfo {pages} {10926} (\bibinfo {year}
  {2007})}\BibitemShut {NoStop}%
\bibitem [{\citenamefont {Miles}\ \emph {et~al.}(2011)\citenamefont {Miles},
  \citenamefont {Lott}, \citenamefont {Luengo~Hendriks}, \citenamefont
  {Ludwig}, \citenamefont {Manu}, \citenamefont {Williams},\ and\ \citenamefont
  {Kreitman}}]{miles+al_11}%
  \BibitemOpen
  \bibfield  {author} {\bibinfo {author} {\bibfnamefont {C.~M.}\ \bibnamefont
  {Miles}}, \bibinfo {author} {\bibfnamefont {S.~E.}\ \bibnamefont {Lott}},
  \bibinfo {author} {\bibfnamefont {C.~L.}\ \bibnamefont {Luengo~Hendriks}},
  \bibinfo {author} {\bibfnamefont {M.~Z.}\ \bibnamefont {Ludwig}}, \bibinfo
  {author} {\bibnamefont {Manu}}, \bibinfo {author} {\bibfnamefont {C.~L.}\
  \bibnamefont {Williams}},\ and\ \bibinfo {author} {\bibfnamefont
  {M.}~\bibnamefont {Kreitman}},\ }\bibfield  {title} {\bibinfo {title}
  {{Artificial selection on egg size perturbs early pattern formation in {\em
  Drosophila melanogaster}}},\ }\href@noop {} {\bibfield  {journal} {\bibinfo
  {journal} {Evolution}\ }\textbf {\bibinfo {volume} {65}},\ \bibinfo {pages}
  {33} (\bibinfo {year} {2011})}\BibitemShut {NoStop}%
\bibitem [{\citenamefont {Antonetti}\ \emph {et~al.}(2018)\citenamefont
  {Antonetti}, \citenamefont {Bialek}, \citenamefont {Gregor}, \citenamefont
  {Muhaxheri}, \citenamefont {Petkova},\ and\ \citenamefont
  {Scheeler}}]{antonetti+al_18}%
  \BibitemOpen
  \bibfield  {author} {\bibinfo {author} {\bibfnamefont {V.}~\bibnamefont
  {Antonetti}}, \bibinfo {author} {\bibfnamefont {W.}~\bibnamefont {Bialek}},
  \bibinfo {author} {\bibfnamefont {T.}~\bibnamefont {Gregor}}, \bibinfo
  {author} {\bibfnamefont {G.}~\bibnamefont {Muhaxheri}}, \bibinfo {author}
  {\bibfnamefont {M.}~\bibnamefont {Petkova}},\ and\ \bibinfo {author}
  {\bibfnamefont {M.}~\bibnamefont {Scheeler}},\ }\bibfield  {title} {\bibinfo
  {title} {Precise spatial scaling in the early fly embryo},\ }\href@noop {}
  {\bibfield  {journal} {\bibinfo  {journal} {arXiv:1812.11384}\ } (\bibinfo
  {year} {2018})}\BibitemShut {NoStop}%
\bibitem [{\citenamefont {Driever}\ and\ \citenamefont
  {N{\"u}sslein-Volhard}(1988{\natexlab{a}})}]{driever+nusslein-volhard_88a}%
  \BibitemOpen
  \bibfield  {author} {\bibinfo {author} {\bibfnamefont {W.}~\bibnamefont
  {Driever}}\ and\ \bibinfo {author} {\bibfnamefont {C.}~\bibnamefont
  {N{\"u}sslein-Volhard}},\ }\bibfield  {title} {\bibinfo {title} {{A gradient
  of bicoid protein in {\em Drosophila} embryos}},\ }\href@noop {} {\bibfield
  {journal} {\bibinfo  {journal} {Cell}\ }\textbf {\bibinfo {volume} {54}},\
  \bibinfo {pages} {83} (\bibinfo {year} {1988}{\natexlab{a}})}\BibitemShut
  {NoStop}%
\bibitem [{\citenamefont {Driever}\ and\ \citenamefont
  {N{\"u}sslein-Volhard}(1988{\natexlab{b}})}]{driever+nusslein-volhard_88b}%
  \BibitemOpen
  \bibfield  {author} {\bibinfo {author} {\bibfnamefont {W.}~\bibnamefont
  {Driever}}\ and\ \bibinfo {author} {\bibfnamefont {C.}~\bibnamefont
  {N{\"u}sslein-Volhard}},\ }\bibfield  {title} {\bibinfo {title} {{The bicoid
  protein determines position in the {\em Drosophila} embryo in a
  concentration-dependent manner}},\ }\href
  {https://doi.org/https://doi.org/10.1016/0092-8674(88)90183-3} {\bibfield
  {journal} {\bibinfo  {journal} {Cell}\ }\textbf {\bibinfo {volume} {54}},\
  \bibinfo {pages} {95} (\bibinfo {year} {1988}{\natexlab{b}})}\BibitemShut
  {NoStop}%
\bibitem [{\citenamefont {Krotov}\ \emph {et~al.}(2014)\citenamefont {Krotov},
  \citenamefont {Dubuis}, \citenamefont {Gregor},\ and\ \citenamefont
  {Bialek}}]{krotov+al_14}%
  \BibitemOpen
  \bibfield  {author} {\bibinfo {author} {\bibfnamefont {D.}~\bibnamefont
  {Krotov}}, \bibinfo {author} {\bibfnamefont {J.~O.}\ \bibnamefont {Dubuis}},
  \bibinfo {author} {\bibfnamefont {T.}~\bibnamefont {Gregor}},\ and\ \bibinfo
  {author} {\bibfnamefont {W.}~\bibnamefont {Bialek}},\ }\bibfield  {title}
  {\bibinfo {title} {Morphogenesis at criticality},\ }\href@noop {} {\bibfield
  {journal} {\bibinfo  {journal} {Proceedings of the National Academy of
  Sciences (USA)}\ }\textbf {\bibinfo {volume} {111}},\ \bibinfo {pages} {3683}
  (\bibinfo {year} {2014})}\BibitemShut {NoStop}%
\bibitem [{\citenamefont {McGough}\ \emph {et~al.}(2023)\citenamefont
  {McGough}, \citenamefont {Casademunt}, \citenamefont {Nikoli\'c},
  \citenamefont {Petkova}, \citenamefont {Gergor},\ and\ \citenamefont
  {Bialek}}]{mcgough+al_23}%
  \BibitemOpen
  \bibfield  {author} {\bibinfo {author} {\bibfnamefont {L.}~\bibnamefont
  {McGough}}, \bibinfo {author} {\bibfnamefont {H.}~\bibnamefont {Casademunt}},
  \bibinfo {author} {\bibfnamefont {M.}~\bibnamefont {Nikoli\'c}}, \bibinfo
  {author} {\bibfnamefont {M.}~\bibnamefont {Petkova}}, \bibinfo {author}
  {\bibfnamefont {T.}~\bibnamefont {Gergor}},\ and\ \bibinfo {author}
  {\bibfnamefont {W.}~\bibnamefont {Bialek}},\ }\bibfield  {title} {\bibinfo
  {title} {Finding the last bits of positional information},\ }\href@noop {}
  {\bibfield  {journal} {\bibinfo  {journal} {arXiv:2312.05963}\ } (\bibinfo
  {year} {2023})}\BibitemShut {NoStop}%
\bibitem [{\citenamefont {Manu}\ \emph {et~al.}(2009)\citenamefont {Manu},
  \citenamefont {Surkova}, \citenamefont {Spirov}, \citenamefont {Gurksy},
  \citenamefont {Janssens}, \citenamefont {Ah-Ram}, \citenamefont {Radulescu},
  \citenamefont {Vanario-Alonso}, \citenamefont {Sharp}, \citenamefont
  {Samsonova},\ and\ \citenamefont {Reinitz}}]{manu+al_09b}%
  \BibitemOpen
  \bibfield  {author} {\bibinfo {author} {\bibnamefont {Manu}}, \bibinfo
  {author} {\bibfnamefont {S.}~\bibnamefont {Surkova}}, \bibinfo {author}
  {\bibfnamefont {A.~V.}\ \bibnamefont {Spirov}}, \bibinfo {author}
  {\bibfnamefont {V.~V.}\ \bibnamefont {Gurksy}}, \bibinfo {author}
  {\bibfnamefont {H.}~\bibnamefont {Janssens}}, \bibinfo {author}
  {\bibfnamefont {K.}~\bibnamefont {Ah-Ram}}, \bibinfo {author} {\bibfnamefont
  {O.}~\bibnamefont {Radulescu}}, \bibinfo {author} {\bibfnamefont {C.~E.}\
  \bibnamefont {Vanario-Alonso}}, \bibinfo {author} {\bibfnamefont {D.~H.}\
  \bibnamefont {Sharp}}, \bibinfo {author} {\bibfnamefont {M.}~\bibnamefont
  {Samsonova}},\ and\ \bibinfo {author} {\bibfnamefont {J.}~\bibnamefont
  {Reinitz}},\ }\bibfield  {title} {\bibinfo {title} {{Canalization of gene
  expression and domain shifts in the {\em Drosophila} blastoderm by dynamical
  attractors}},\ }\href@noop {} {\bibfield  {journal} {\bibinfo  {journal}
  {PLoS Computational Biology}\ }\textbf {\bibinfo {volume} {5}},\ \bibinfo
  {pages} {e1000303} (\bibinfo {year} {2009})}\BibitemShut {NoStop}%
\bibitem [{\citenamefont {Vakulenko}\ \emph {et~al.}(2009)\citenamefont
  {Vakulenko}, \citenamefont {Manu}, \citenamefont {Reinitz},\ and\
  \citenamefont {Radulescu}}]{vakulenko+al_09}%
  \BibitemOpen
  \bibfield  {author} {\bibinfo {author} {\bibfnamefont {S.}~\bibnamefont
  {Vakulenko}}, \bibinfo {author} {\bibnamefont {Manu}}, \bibinfo {author}
  {\bibfnamefont {J.}~\bibnamefont {Reinitz}},\ and\ \bibinfo {author}
  {\bibfnamefont {O.}~\bibnamefont {Radulescu}},\ }\bibfield  {title} {\bibinfo
  {title} {{Size regulation in the segmentation of {\em Drosophila}:
  Interacting interfaces between localized domains of gene expression ensure
  robust spatial patterning}},\ }\href@noop {} {\bibfield  {journal} {\bibinfo
  {journal} {Physical Review Letters}\ }\textbf {\bibinfo {volume} {103}},\
  \bibinfo {pages} {168102} (\bibinfo {year} {2009})}\BibitemShut {NoStop}%
\bibitem [{\citenamefont {Wolpert}(1969)}]{wolpert_69}%
  \BibitemOpen
  \bibfield  {author} {\bibinfo {author} {\bibfnamefont {L.}~\bibnamefont
  {Wolpert}},\ }\bibfield  {title} {\bibinfo {title} {Positional information
  and the spatial pattern of cellular differentiation},\ }\href@noop {}
  {\bibfield  {journal} {\bibinfo  {journal} {Journal of Theoretical Biology}\
  }\textbf {\bibinfo {volume} {25}},\ \bibinfo {pages} {1} (\bibinfo {year}
  {1969})}\BibitemShut {NoStop}%
\bibitem [{\citenamefont {Lawrence}(1992)}]{lawrence_92}%
  \BibitemOpen
  \bibfield  {author} {\bibinfo {author} {\bibfnamefont {P.~A.}\ \bibnamefont
  {Lawrence}},\ }\href@noop {} {\emph {\bibinfo {title} {The Making of a Fly:
  The Genetics of Animal Design}}}\ (\bibinfo  {publisher} {Blackwell
  Scientific Publications Ltd},\ \bibinfo {address} {Oxford},\ \bibinfo {year}
  {1992})\BibitemShut {NoStop}%
\bibitem [{\citenamefont {Dubuis}\ \emph
  {et~al.}(2013{\natexlab{a}})\citenamefont {Dubuis}, \citenamefont {Samanta},\
  and\ \citenamefont {Gregor}}]{dubuis+al_13a}%
  \BibitemOpen
  \bibfield  {author} {\bibinfo {author} {\bibfnamefont {J.~O.}\ \bibnamefont
  {Dubuis}}, \bibinfo {author} {\bibfnamefont {R.}~\bibnamefont {Samanta}},\
  and\ \bibinfo {author} {\bibfnamefont {T.}~\bibnamefont {Gregor}},\
  }\bibfield  {title} {\bibinfo {title} {Accurate measurements of dynamics and
  reproducibility in small genetic networks},\ }\href@noop {} {\bibfield
  {journal} {\bibinfo  {journal} {Molecular Systems Biology}\ }\textbf
  {\bibinfo {volume} {9}},\ \bibinfo {pages} {639} (\bibinfo {year}
  {2013}{\natexlab{a}})}\BibitemShut {NoStop}%
\bibitem [{\citenamefont {Petkova}\ \emph {et~al.}(2019)\citenamefont
  {Petkova}, \citenamefont {Tka{\v{c}}ik}, \citenamefont {Bialek},
  \citenamefont {Wieschaus},\ and\ \citenamefont {Gregor}}]{petkova+al_19}%
  \BibitemOpen
  \bibfield  {author} {\bibinfo {author} {\bibfnamefont {M.~D.}\ \bibnamefont
  {Petkova}}, \bibinfo {author} {\bibfnamefont {G.}~\bibnamefont
  {Tka{\v{c}}ik}}, \bibinfo {author} {\bibfnamefont {W.}~\bibnamefont
  {Bialek}}, \bibinfo {author} {\bibfnamefont {E.~F.}\ \bibnamefont
  {Wieschaus}},\ and\ \bibinfo {author} {\bibfnamefont {T.}~\bibnamefont
  {Gregor}},\ }\bibfield  {title} {\bibinfo {title} {Optimal decoding of
  cellular identities in a genetic network},\ }\href@noop {} {\bibfield
  {journal} {\bibinfo  {journal} {Cell}\ }\textbf {\bibinfo {volume} {176}},\
  \bibinfo {pages} {844} (\bibinfo {year} {2019})}\BibitemShut {NoStop}%
\bibitem [{\citenamefont {Gregor}\ \emph
  {et~al.}(2007{\natexlab{a}})\citenamefont {Gregor}, \citenamefont
  {Wieschaus}, \citenamefont {McGregor}, \citenamefont {Bialek},\ and\
  \citenamefont {Tank}}]{gregor+al_07a}%
  \BibitemOpen
  \bibfield  {author} {\bibinfo {author} {\bibfnamefont {T.}~\bibnamefont
  {Gregor}}, \bibinfo {author} {\bibfnamefont {E.~F.}\ \bibnamefont
  {Wieschaus}}, \bibinfo {author} {\bibfnamefont {A.~P.}\ \bibnamefont
  {McGregor}}, \bibinfo {author} {\bibfnamefont {W.}~\bibnamefont {Bialek}},\
  and\ \bibinfo {author} {\bibfnamefont {D.~W.}\ \bibnamefont {Tank}},\
  }\bibfield  {title} {\bibinfo {title} {Stability and nuclear dynamics of the
  bicoid morphogen gradient},\ }\href@noop {} {\bibfield  {journal} {\bibinfo
  {journal} {Cell}\ }\textbf {\bibinfo {volume} {130}},\ \bibinfo {pages} {141}
  (\bibinfo {year} {2007}{\natexlab{a}})}\BibitemShut {NoStop}%
\bibitem [{\citenamefont {Smith}(2015)}]{smith_15}%
  \BibitemOpen
  \bibfield  {author} {\bibinfo {author} {\bibfnamefont {E.~M.}\ \bibnamefont
  {Smith}},\ }\emph {\bibinfo {title} {Scaling and Regulation of Gene
  Expression in the Developing Fly Embryo}},\ \href@noop {} {Ph.D. thesis},\
  \bibinfo  {school} {Princeton University} (\bibinfo {year}
  {2015})\BibitemShut {NoStop}%
\bibitem [{\citenamefont {Shannon}(1948)}]{shannon_48}%
  \BibitemOpen
  \bibfield  {author} {\bibinfo {author} {\bibfnamefont {C.~E.}\ \bibnamefont
  {Shannon}},\ }\bibfield  {title} {\bibinfo {title} {A mathematical theory of
  communication},\ }\href@noop {} {\bibfield  {journal} {\bibinfo  {journal}
  {Bell System Technical Journal}\ }\textbf {\bibinfo {volume} {27}},\ \bibinfo
  {pages} {379} (\bibinfo {year} {1948})}\BibitemShut {NoStop}%
\bibitem [{\citenamefont {Cover}\ and\ \citenamefont
  {Thomas}(1991)}]{cover+thomas_91}%
  \BibitemOpen
  \bibfield  {author} {\bibinfo {author} {\bibfnamefont {T.~M.}\ \bibnamefont
  {Cover}}\ and\ \bibinfo {author} {\bibfnamefont {J.~A.}\ \bibnamefont
  {Thomas}},\ }\href@noop {} {\emph {\bibinfo {title} {Elements of Information
  Theory}}}\ (\bibinfo  {publisher} {Wiley and Sons},\ \bibinfo {address} {New
  York},\ \bibinfo {year} {1991})\BibitemShut {NoStop}%
\bibitem [{\citenamefont {Bialek}(2012)}]{bialek_12}%
  \BibitemOpen
  \bibfield  {author} {\bibinfo {author} {\bibfnamefont {W.}~\bibnamefont
  {Bialek}},\ }\href@noop {} {\emph {\bibinfo {title} {Biophysics: Searching
  for Principles}}}\ (\bibinfo  {publisher} {Princeton University Press},\
  \bibinfo {year} {2012})\BibitemShut {NoStop}%
\bibitem [{\citenamefont {Howard}\ and\ \citenamefont {ten
  Wolde}(2005)}]{howard+wolde_05}%
  \BibitemOpen
  \bibfield  {author} {\bibinfo {author} {\bibfnamefont {M.}~\bibnamefont
  {Howard}}\ and\ \bibinfo {author} {\bibfnamefont {P.~R.}\ \bibnamefont {ten
  Wolde}},\ }\bibfield  {title} {\bibinfo {title} {Finding the center reliably:
  Robust patterns of developmental gene expression},\ }\href@noop {} {\bibfield
   {journal} {\bibinfo  {journal} {Physical review letters}\ }\textbf {\bibinfo
  {volume} {95}},\ \bibinfo {pages} {208103} (\bibinfo {year}
  {2005})}\BibitemShut {NoStop}%
\bibitem [{\citenamefont {Houchmandzadeh}\ \emph {et~al.}(2005)\citenamefont
  {Houchmandzadeh}, \citenamefont {Wieschaus},\ and\ \citenamefont
  {Leibler}}]{houchmandzadeh+al_05}%
  \BibitemOpen
  \bibfield  {author} {\bibinfo {author} {\bibfnamefont {B.}~\bibnamefont
  {Houchmandzadeh}}, \bibinfo {author} {\bibfnamefont {E.}~\bibnamefont
  {Wieschaus}},\ and\ \bibinfo {author} {\bibfnamefont {S.}~\bibnamefont
  {Leibler}},\ }\bibfield  {title} {\bibinfo {title} {{Precise domain
  specification in the developing {\em Drosophila} embryo}},\ }\href@noop {}
  {\bibfield  {journal} {\bibinfo  {journal} {Physical Review E}\ }\textbf
  {\bibinfo {volume} {72}},\ \bibinfo {pages} {061920} (\bibinfo {year}
  {2005})}\BibitemShut {NoStop}%
\bibitem [{\citenamefont {McHale}\ \emph {et~al.}(2006)\citenamefont {McHale},
  \citenamefont {Rappel},\ and\ \citenamefont {Levine}}]{mchale+al_06}%
  \BibitemOpen
  \bibfield  {author} {\bibinfo {author} {\bibfnamefont {P.}~\bibnamefont
  {McHale}}, \bibinfo {author} {\bibfnamefont {W.-J.}\ \bibnamefont {Rappel}},\
  and\ \bibinfo {author} {\bibfnamefont {H.}~\bibnamefont {Levine}},\
  }\bibfield  {title} {\bibinfo {title} {Embryonic pattern scaling achieved by
  oppositely directed morphogen gradients},\ }\href@noop {} {\bibfield
  {journal} {\bibinfo  {journal} {Physical Biology}\ }\textbf {\bibinfo
  {volume} {3}},\ \bibinfo {pages} {107} (\bibinfo {year} {2006})}\BibitemShut
  {NoStop}%
\bibitem [{\citenamefont {\v{C}apek}\ and\ \citenamefont
  {M\"uller}(2019)}]{capek+muller_19}%
  \BibitemOpen
  \bibfield  {author} {\bibinfo {author} {\bibfnamefont {D.}~\bibnamefont
  {\v{C}apek}}\ and\ \bibinfo {author} {\bibfnamefont {P.}~\bibnamefont
  {M\"uller}},\ }\bibfield  {title} {\bibinfo {title} {Positional information
  and tissue scaling during development and regeneration},\ }\href@noop {}
  {\bibfield  {journal} {\bibinfo  {journal} {Development}\ }\textbf {\bibinfo
  {volume} {146}},\ \bibinfo {pages} {dev177709} (\bibinfo {year}
  {2019})}\BibitemShut {NoStop}%
\bibitem [{\citenamefont {Frasch}\ \emph {et~al.}(1988)\citenamefont {Frasch},
  \citenamefont {Warrior}, \citenamefont {Tugwood},\ and\ \citenamefont
  {Levine}}]{frasch+al_88}%
  \BibitemOpen
  \bibfield  {author} {\bibinfo {author} {\bibfnamefont {M.}~\bibnamefont
  {Frasch}}, \bibinfo {author} {\bibfnamefont {R.}~\bibnamefont {Warrior}},
  \bibinfo {author} {\bibfnamefont {J.}~\bibnamefont {Tugwood}},\ and\ \bibinfo
  {author} {\bibfnamefont {M.}~\bibnamefont {Levine}},\ }\bibfield  {title}
  {\bibinfo {title} {{Molecular analysis of even-skipped mutants in {\em
  Drosophila} development}},\ }\href@noop {} {\bibfield  {journal} {\bibinfo
  {journal} {Genes and Development}\ }\textbf {\bibinfo {volume} {2}},\
  \bibinfo {pages} {1824} (\bibinfo {year} {1988})}\BibitemShut {NoStop}%
\bibitem [{\citenamefont {Jaeger}\ \emph
  {et~al.}(2004{\natexlab{a}})\citenamefont {Jaeger}, \citenamefont {Surkova},
  \citenamefont {Blagov}, \citenamefont {Janssens}, \citenamefont {Kosman},
  \citenamefont {Kozlov}, \citenamefont {Myasnikova}, \citenamefont
  {Vanario-Alonso}, \citenamefont {Samsonova}, \citenamefont {Sharp},\ and\
  \citenamefont {Reinitz}}]{jaeger+al_04a}%
  \BibitemOpen
  \bibfield  {author} {\bibinfo {author} {\bibfnamefont {J.}~\bibnamefont
  {Jaeger}}, \bibinfo {author} {\bibfnamefont {S.}~\bibnamefont {Surkova}},
  \bibinfo {author} {\bibfnamefont {M.}~\bibnamefont {Blagov}}, \bibinfo
  {author} {\bibfnamefont {H.}~\bibnamefont {Janssens}}, \bibinfo {author}
  {\bibfnamefont {D.}~\bibnamefont {Kosman}}, \bibinfo {author} {\bibfnamefont
  {K.~N.}\ \bibnamefont {Kozlov}}, \bibinfo {author} {\bibfnamefont
  {E.}~\bibnamefont {Myasnikova}}, \bibinfo {author} {\bibfnamefont {C.~E.}\
  \bibnamefont {Vanario-Alonso}}, \bibinfo {author} {\bibfnamefont
  {M.}~\bibnamefont {Samsonova}}, \bibinfo {author} {\bibfnamefont {D.~H.}\
  \bibnamefont {Sharp}},\ and\ \bibinfo {author} {\bibfnamefont
  {J.}~\bibnamefont {Reinitz}},\ }\bibfield  {title} {\bibinfo {title}
  {{Dynamic control of positional information in the early {\em Drosophila}
  embryo}},\ }\href@noop {} {\bibfield  {journal} {\bibinfo  {journal}
  {Nature}\ }\textbf {\bibinfo {volume} {430}},\ \bibinfo {pages} {368}
  (\bibinfo {year} {2004}{\natexlab{a}})}\BibitemShut {NoStop}%
\bibitem [{\citenamefont {Jaeger}\ \emph
  {et~al.}(2004{\natexlab{b}})\citenamefont {Jaeger}, \citenamefont {Blagov},
  \citenamefont {Kosman}, \citenamefont {Kozlov}, \citenamefont {Manu},
  \citenamefont {Myasnikova}, \citenamefont {Surkova}, \citenamefont
  {Vanario-Alonso}, \citenamefont {Samsonova}, \citenamefont {Sharp},\ and\
  \citenamefont {Reinitz}}]{jaeger+al_04b}%
  \BibitemOpen
  \bibfield  {author} {\bibinfo {author} {\bibfnamefont {J.}~\bibnamefont
  {Jaeger}}, \bibinfo {author} {\bibfnamefont {M.}~\bibnamefont {Blagov}},
  \bibinfo {author} {\bibfnamefont {D.}~\bibnamefont {Kosman}}, \bibinfo
  {author} {\bibfnamefont {K.~N.}\ \bibnamefont {Kozlov}}, \bibinfo {author}
  {\bibnamefont {Manu}}, \bibinfo {author} {\bibfnamefont {E.}~\bibnamefont
  {Myasnikova}}, \bibinfo {author} {\bibfnamefont {S.}~\bibnamefont {Surkova}},
  \bibinfo {author} {\bibfnamefont {C.~E.}\ \bibnamefont {Vanario-Alonso}},
  \bibinfo {author} {\bibfnamefont {M.}~\bibnamefont {Samsonova}}, \bibinfo
  {author} {\bibfnamefont {D.~H.}\ \bibnamefont {Sharp}},\ and\ \bibinfo
  {author} {\bibfnamefont {J.}~\bibnamefont {Reinitz}},\ }\bibfield  {title}
  {\bibinfo {title} {{Dynamical analysis of regulatory interactions in the gap
  gene system of {\em Drosophila melanogaster}}},\ }\href@noop {} {\bibfield
  {journal} {\bibinfo  {journal} {Genetics}\ }\textbf {\bibinfo {volume}
  {167}},\ \bibinfo {pages} {1721} (\bibinfo {year}
  {2004}{\natexlab{b}})}\BibitemShut {NoStop}%
\bibitem [{\citenamefont {Bothma}\ \emph {et~al.}(2014)\citenamefont {Bothma},
  \citenamefont {Garcia}, \citenamefont {Esposito}, \citenamefont {Schlissel},
  \citenamefont {Gregor},\ and\ \citenamefont {Levine}}]{bothma+al_14}%
  \BibitemOpen
  \bibfield  {author} {\bibinfo {author} {\bibfnamefont {J.~P.}\ \bibnamefont
  {Bothma}}, \bibinfo {author} {\bibfnamefont {H.~G.}\ \bibnamefont {Garcia}},
  \bibinfo {author} {\bibfnamefont {E.}~\bibnamefont {Esposito}}, \bibinfo
  {author} {\bibfnamefont {G.}~\bibnamefont {Schlissel}}, \bibinfo {author}
  {\bibfnamefont {T.}~\bibnamefont {Gregor}},\ and\ \bibinfo {author}
  {\bibfnamefont {M.}~\bibnamefont {Levine}},\ }\bibfield  {title} {\bibinfo
  {title} {{Dynamic regulation of eve stripe 2 expression reveals
  transcriptional bursts in living {\em Drosophila} embryos}},\ }\href@noop {}
  {\bibfield  {journal} {\bibinfo  {journal} {Proceedings of the National
  Academy of Sciences (USA)}\ }\textbf {\bibinfo {volume} {111}},\ \bibinfo
  {pages} {10598} (\bibinfo {year} {2014})}\BibitemShut {NoStop}%
\bibitem [{\citenamefont {Dubuis}\ \emph
  {et~al.}(2013{\natexlab{b}})\citenamefont {Dubuis}, \citenamefont
  {Tka{\v{c}}ik}, \citenamefont {Wieschaus}, \citenamefont {Gregor},\ and\
  \citenamefont {Bialek}}]{dubuis+al_13b}%
  \BibitemOpen
  \bibfield  {author} {\bibinfo {author} {\bibfnamefont {J.~O.}\ \bibnamefont
  {Dubuis}}, \bibinfo {author} {\bibfnamefont {G.}~\bibnamefont
  {Tka{\v{c}}ik}}, \bibinfo {author} {\bibfnamefont {E.~F.}\ \bibnamefont
  {Wieschaus}}, \bibinfo {author} {\bibfnamefont {T.}~\bibnamefont {Gregor}},\
  and\ \bibinfo {author} {\bibfnamefont {W.}~\bibnamefont {Bialek}},\
  }\bibfield  {title} {\bibinfo {title} {Positional information, in bits},\
  }\href@noop {} {\bibfield  {journal} {\bibinfo  {journal} {Proceedings of the
  National Academy of Sciences (USA)}\ }\textbf {\bibinfo {volume} {110}},\
  \bibinfo {pages} {16301} (\bibinfo {year} {2013}{\natexlab{b}})}\BibitemShut
  {NoStop}%
\bibitem [{\citenamefont {Vincent}\ \emph {et~al.}(1997)\citenamefont
  {Vincent}, \citenamefont {Blankenship},\ and\ \citenamefont
  {Wieschaus}}]{vincent+al_97}%
  \BibitemOpen
  \bibfield  {author} {\bibinfo {author} {\bibfnamefont {A.}~\bibnamefont
  {Vincent}}, \bibinfo {author} {\bibfnamefont {J.~T.}\ \bibnamefont
  {Blankenship}},\ and\ \bibinfo {author} {\bibfnamefont {E.}~\bibnamefont
  {Wieschaus}},\ }\bibfield  {title} {\bibinfo {title} {{Integration of the
  head and trunk segmentation systems controls cephalic furrow formation in
  {\em Drosophila}}},\ }\href@noop {} {\bibfield  {journal} {\bibinfo
  {journal} {Development}\ }\textbf {\bibinfo {volume} {124}},\ \bibinfo
  {pages} {3747} (\bibinfo {year} {1997})}\BibitemShut {NoStop}%
\bibitem [{\citenamefont {Liu}\ \emph {et~al.}(2013)\citenamefont {Liu},
  \citenamefont {Morrison},\ and\ \citenamefont {Gregor}}]{liu+al_13}%
  \BibitemOpen
  \bibfield  {author} {\bibinfo {author} {\bibfnamefont {F.}~\bibnamefont
  {Liu}}, \bibinfo {author} {\bibfnamefont {A.~H.}\ \bibnamefont {Morrison}},\
  and\ \bibinfo {author} {\bibfnamefont {T.}~\bibnamefont {Gregor}},\
  }\bibfield  {title} {\bibinfo {title} {{Dynamic interpretation of maternal
  inputs by the {\em Drosophila} segmentation gene network}},\ }\href@noop {}
  {\bibfield  {journal} {\bibinfo  {journal} {Proceedings of the National
  Academy of Sciences (USA)}\ }\textbf {\bibinfo {volume} {110}},\ \bibinfo
  {pages} {6724} (\bibinfo {year} {2013})}\BibitemShut {NoStop}%
\bibitem [{\citenamefont {Jaeger}(2011)}]{jaeger_11}%
  \BibitemOpen
  \bibfield  {author} {\bibinfo {author} {\bibfnamefont {J.}~\bibnamefont
  {Jaeger}},\ }\bibfield  {title} {\bibinfo {title} {The gap gene network},\
  }\href@noop {} {\bibfield  {journal} {\bibinfo  {journal} {Cellular and
  Molecular Life Sciences}\ }\textbf {\bibinfo {volume} {68}},\ \bibinfo
  {pages} {243} (\bibinfo {year} {2011})}\BibitemShut {NoStop}%
\bibitem [{\citenamefont {Kauffman}\ \emph {et~al.}(1978)\citenamefont
  {Kauffman}, \citenamefont {Shymko},\ and\ \citenamefont
  {Trabert}}]{kauffman+al_78}%
  \BibitemOpen
  \bibfield  {author} {\bibinfo {author} {\bibfnamefont {S.~A.}\ \bibnamefont
  {Kauffman}}, \bibinfo {author} {\bibfnamefont {R.~M.}\ \bibnamefont
  {Shymko}},\ and\ \bibinfo {author} {\bibfnamefont {K.}~\bibnamefont
  {Trabert}},\ }\bibfield  {title} {\bibinfo {title} {{Control of sequential
  compartment formation in {\em Drosophila}: A uniform mechanism may control
  the locations of successive binary developmental commitments}},\ }\href@noop
  {} {\bibfield  {journal} {\bibinfo  {journal} {Science}\ }\textbf {\bibinfo
  {volume} {199}},\ \bibinfo {pages} {259} (\bibinfo {year}
  {1978})}\BibitemShut {NoStop}%
\bibitem [{\citenamefont {Meinhardt}(1986)}]{meinhardt_86}%
  \BibitemOpen
  \bibfield  {author} {\bibinfo {author} {\bibfnamefont {H.}~\bibnamefont
  {Meinhardt}},\ }\bibfield  {title} {\bibinfo {title} {{Hierarchical
  inductions of cell states: A model for segmentation in {\em Drosophila}}},\
  }\href@noop {} {\bibfield  {journal} {\bibinfo  {journal} {Journal of Cell
  Science Supplement}\ }\textbf {\bibinfo {volume} {4}},\ \bibinfo {pages}
  {357} (\bibinfo {year} {1986})}\BibitemShut {NoStop}%
\bibitem [{\citenamefont {Albert}\ and\ \citenamefont
  {Othmer}(2003)}]{albert+othmer_03}%
  \BibitemOpen
  \bibfield  {author} {\bibinfo {author} {\bibfnamefont {R.}~\bibnamefont
  {Albert}}\ and\ \bibinfo {author} {\bibfnamefont {H.~G.}\ \bibnamefont
  {Othmer}},\ }\bibfield  {title} {\bibinfo {title} {{The topology of the
  regulatory interactions predicts the expression pattern of the segment
  polarity genes in {\em Drosophila} melanogaster}},\ }\href@noop {} {\bibfield
   {journal} {\bibinfo  {journal} {Journal of Theoretical Biology}\ }\textbf
  {\bibinfo {volume} {223}},\ \bibinfo {pages} {1} (\bibinfo {year}
  {2003})}\BibitemShut {NoStop}%
\bibitem [{\citenamefont {Spirov}\ and\ \citenamefont
  {Holloway}(2003)}]{spirov+holloway_86}%
  \BibitemOpen
  \bibfield  {author} {\bibinfo {author} {\bibfnamefont {A.~V.}\ \bibnamefont
  {Spirov}}\ and\ \bibinfo {author} {\bibfnamefont {D.~M.}\ \bibnamefont
  {Holloway}},\ }\bibfield  {title} {\bibinfo {title} {{Making the body plan:
  Precision in the genetic hierarchy of {\em Drosophila} embryo
  segmentation}},\ }\href@noop {} {\bibfield  {journal} {\bibinfo  {journal}
  {In Silico Biology}\ }\textbf {\bibinfo {volume} {3}},\ \bibinfo {pages} {89}
  (\bibinfo {year} {2003})}\BibitemShut {NoStop}%
\bibitem [{\citenamefont {Tka{\v{c}}ik}\ \emph {et~al.}(2015)\citenamefont
  {Tka{\v{c}}ik}, \citenamefont {Dubuis}, \citenamefont {Petkova},\ and\
  \citenamefont {Gregor}}]{tkacik+al_15}%
  \BibitemOpen
  \bibfield  {author} {\bibinfo {author} {\bibfnamefont {G.}~\bibnamefont
  {Tka{\v{c}}ik}}, \bibinfo {author} {\bibfnamefont {J.~O.}\ \bibnamefont
  {Dubuis}}, \bibinfo {author} {\bibfnamefont {M.~D.}\ \bibnamefont
  {Petkova}},\ and\ \bibinfo {author} {\bibfnamefont {T.}~\bibnamefont
  {Gregor}},\ }\bibfield  {title} {\bibinfo {title} {Positional information,
  positional error, and readout precision in morphogenesis: A mathematical
  framework},\ }\href {https://doi.org/10.1534/genetics.114.171850} {\bibfield
  {journal} {\bibinfo  {journal} {Genetics}\ }\textbf {\bibinfo {volume}
  {199}},\ \bibinfo {pages} {39} (\bibinfo {year} {2015})}\BibitemShut
  {NoStop}%
\bibitem [{\citenamefont {Miller}(1955)}]{miller_55}%
  \BibitemOpen
  \bibfield  {author} {\bibinfo {author} {\bibfnamefont {G.}~\bibnamefont
  {Miller}},\ }\bibfield  {title} {\bibinfo {title} {Note on the bias of
  information estimates},\ }in\ \href@noop {} {\emph {\bibinfo {booktitle}
  {Information Theory in Psychology II--B: Problems and Methods}}},\ \bibinfo
  {editor} {edited by\ \bibinfo {editor} {\bibfnamefont {H.}~\bibnamefont
  {Quastler}}}\ (\bibinfo  {publisher} {Free Press},\ \bibinfo {address}
  {Glencoe IL},\ \bibinfo {year} {1955})\ pp.\ \bibinfo {pages}
  {95--100}\BibitemShut {NoStop}%
\bibitem [{\citenamefont {Gregor}\ \emph {et~al.}(2005)\citenamefont {Gregor},
  \citenamefont {Bialek}, \citenamefont {de~Ruyter~van Steveninck},
  \citenamefont {Tank},\ and\ \citenamefont {Wieschaus}}]{gregor+al_05}%
  \BibitemOpen
  \bibfield  {author} {\bibinfo {author} {\bibfnamefont {T.}~\bibnamefont
  {Gregor}}, \bibinfo {author} {\bibfnamefont {W.}~\bibnamefont {Bialek}},
  \bibinfo {author} {\bibfnamefont {R.~R.}\ \bibnamefont {de~Ruyter~van
  Steveninck}}, \bibinfo {author} {\bibfnamefont {D.~W.}\ \bibnamefont
  {Tank}},\ and\ \bibinfo {author} {\bibfnamefont {E.~F.}\ \bibnamefont
  {Wieschaus}},\ }\bibfield  {title} {\bibinfo {title} {Diffusion and scaling
  during early embryonic pattern formation},\ }\href@noop {} {\bibfield
  {journal} {\bibinfo  {journal} {Proceedings of the National Academy of
  Sciences}\ }\textbf {\bibinfo {volume} {102}},\ \bibinfo {pages} {18403}
  (\bibinfo {year} {2005})}\BibitemShut {NoStop}%
\bibitem [{\citenamefont {Gregor}\ \emph {et~al.}(2008)\citenamefont {Gregor},
  \citenamefont {McGregor},\ and\ \citenamefont {Wieschaus}}]{gregor+al_08}%
  \BibitemOpen
  \bibfield  {author} {\bibinfo {author} {\bibfnamefont {T.}~\bibnamefont
  {Gregor}}, \bibinfo {author} {\bibfnamefont {A.~P.}\ \bibnamefont
  {McGregor}},\ and\ \bibinfo {author} {\bibfnamefont {E.~F.}\ \bibnamefont
  {Wieschaus}},\ }\bibfield  {title} {\bibinfo {title} {Shape and function of
  the bicoid morphogen gradient in dipteran species with different sized
  embryos},\ }\href@noop {} {\bibfield  {journal} {\bibinfo  {journal}
  {Developmental Biology}\ }\textbf {\bibinfo {volume} {316}},\ \bibinfo
  {pages} {350} (\bibinfo {year} {2008})}\BibitemShut {NoStop}%
\bibitem [{\citenamefont {Gregor}\ \emph
  {et~al.}(2007{\natexlab{b}})\citenamefont {Gregor}, \citenamefont {Tank},
  \citenamefont {Wieschaus},\ and\ \citenamefont {Bialek}}]{gregor+al_07b}%
  \BibitemOpen
  \bibfield  {author} {\bibinfo {author} {\bibfnamefont {T.}~\bibnamefont
  {Gregor}}, \bibinfo {author} {\bibfnamefont {D.~W.}\ \bibnamefont {Tank}},
  \bibinfo {author} {\bibfnamefont {E.~F.}\ \bibnamefont {Wieschaus}},\ and\
  \bibinfo {author} {\bibfnamefont {W.}~\bibnamefont {Bialek}},\ }\bibfield
  {title} {\bibinfo {title} {Probing the limits to positional information},\
  }\href@noop {} {\bibfield  {journal} {\bibinfo  {journal} {Cell}\ }\textbf
  {\bibinfo {volume} {130}},\ \bibinfo {pages} {153} (\bibinfo {year}
  {2007}{\natexlab{b}})}\BibitemShut {NoStop}%
\bibitem [{\citenamefont {Petkova}\ \emph {et~al.}(2014)\citenamefont
  {Petkova}, \citenamefont {Little}, \citenamefont {Liu},\ and\ \citenamefont
  {Gregor}}]{petkova+al_14}%
  \BibitemOpen
  \bibfield  {author} {\bibinfo {author} {\bibfnamefont {M.~D.}\ \bibnamefont
  {Petkova}}, \bibinfo {author} {\bibfnamefont {S.~C.}\ \bibnamefont {Little}},
  \bibinfo {author} {\bibfnamefont {F.}~\bibnamefont {Liu}},\ and\ \bibinfo
  {author} {\bibfnamefont {T.}~\bibnamefont {Gregor}},\ }\bibfield  {title}
  {\bibinfo {title} {Maternal origins of developmental reproducibility},\
  }\href@noop {} {\bibfield  {journal} {\bibinfo  {journal} {Current Biology}\
  }\textbf {\bibinfo {volume} {24}},\ \bibinfo {pages} {1283} (\bibinfo {year}
  {2014})}\BibitemShut {NoStop}%
\bibitem [{\citenamefont {Gierer}\ and\ \citenamefont
  {Meinhardt}(1972)}]{gierer+meinhardt_72}%
  \BibitemOpen
  \bibfield  {author} {\bibinfo {author} {\bibfnamefont {A.}~\bibnamefont
  {Gierer}}\ and\ \bibinfo {author} {\bibfnamefont {H.}~\bibnamefont
  {Meinhardt}},\ }\bibfield  {title} {\bibinfo {title} {A theory of biological
  pattern formation},\ }\href@noop {} {\bibfield  {journal} {\bibinfo
  {journal} {Kybernetik}\ }\textbf {\bibinfo {volume} {12}},\ \bibinfo {pages}
  {30} (\bibinfo {year} {1972})}\BibitemShut {NoStop}%
\bibitem [{\citenamefont {Meinhardt}(2008)}]{meinhardt_08}%
  \BibitemOpen
  \bibfield  {author} {\bibinfo {author} {\bibfnamefont {H.}~\bibnamefont
  {Meinhardt}},\ }\bibfield  {title} {\bibinfo {title} {Models of biological
  pattern formation: From elementary steps to the organization of embryonic
  axes},\ }\href@noop {} {\bibfield  {journal} {\bibinfo  {journal} {Current
  Topics in Developmental Biology}\ }\textbf {\bibinfo {volume} {81}},\
  \bibinfo {pages} {1} (\bibinfo {year} {2008})}\BibitemShut {NoStop}%
\bibitem [{\citenamefont {Seyboldt}\ \emph {et~al.}(2022)\citenamefont
  {Seyboldt}, \citenamefont {Lavoie}, \citenamefont {Henry},\ and\
  \citenamefont {Fran\c{c}ois}}]{seyboldt+al_22}%
  \BibitemOpen
  \bibfield  {author} {\bibinfo {author} {\bibfnamefont {R.}~\bibnamefont
  {Seyboldt}}, \bibinfo {author} {\bibfnamefont {J.}~\bibnamefont {Lavoie}},
  \bibinfo {author} {\bibfnamefont {A.}~\bibnamefont {Henry}},\ and\ \bibinfo
  {author} {\bibfnamefont {P.}~\bibnamefont {Fran\c{c}ois}},\ }\bibfield
  {title} {\bibinfo {title} {Latent space of a small genetic network: Geometry
  of dynamics and information},\ }\href@noop {} {\bibfield  {journal} {\bibinfo
   {journal} {Proceedings of the National Academy of Sciences (USA)}\ }\textbf
  {\bibinfo {volume} {119}},\ \bibinfo {pages} {e2113651119} (\bibinfo {year}
  {2022})}\BibitemShut {NoStop}%
\bibitem [{\citenamefont {Seung}(1996)}]{seung_96}%
  \BibitemOpen
  \bibfield  {author} {\bibinfo {author} {\bibfnamefont {H.~S.}\ \bibnamefont
  {Seung}},\ }\bibfield  {title} {\bibinfo {title} {How the brain keeps the
  eyes still},\ }\href@noop {} {\bibfield  {journal} {\bibinfo  {journal}
  {Proceedings of the National Academy of Sciences (USA)}\ }\textbf {\bibinfo
  {volume} {93}},\ \bibinfo {pages} {13339} (\bibinfo {year}
  {1996})}\BibitemShut {NoStop}%
\bibitem [{\citenamefont {Merle}\ \emph {et~al.}(2023)\citenamefont {Merle},
  \citenamefont {Friedman}, \citenamefont {Chureau}, \citenamefont
  {Shoushtarizadeh},\ and\ \citenamefont {Gregor}}]{merle+al_23}%
  \BibitemOpen
  \bibfield  {author} {\bibinfo {author} {\bibfnamefont {M.}~\bibnamefont
  {Merle}}, \bibinfo {author} {\bibfnamefont {L.}~\bibnamefont {Friedman}},
  \bibinfo {author} {\bibfnamefont {C.}~\bibnamefont {Chureau}}, \bibinfo
  {author} {\bibfnamefont {A.}~\bibnamefont {Shoushtarizadeh}},\ and\ \bibinfo
  {author} {\bibfnamefont {T.}~\bibnamefont {Gregor}},\ }\bibfield  {title}
  {\bibinfo {title} {Precise and scalable self-organization in mammalian
  pseudo-embryos},\ }\href@noop {} {\bibfield  {journal} {\bibinfo  {journal}
  {arXiv:2303.17522}\ } (\bibinfo {year} {2023})}\BibitemShut {NoStop}%
\end{thebibliography}%

\end{document}